\documentclass[fleqn,10pt]{wlscirep}
\usepackage[utf8]{inputenc}
\usepackage[T1]{fontenc}
\usepackage{multirow}
\title{Robust and Generalizable Atrial Fibrillation Detection from ECG Using Time-Frequency Fusion and Supervised Contrastive Learning}

\author[1]{Hongtao Li}
\author[2]{Jia Wei}
\author[1]{Jia Xiao}
\author[1]{Yuanjun Lai}
\author[2]{Mingyang Liu}
\author[1]{Shuzhen Lv}
\author[1,*]{Xueqiang Ouyang}
\affil[1]{Information Center, The People’s Hospital of Baoan Shenzhen, Shenzhen 518101, Guangdong, China}
\affil[2]{School of Computer Science and Engineering, South China University of Technology, Guangzhou 510006, Guangdong, China}

\affil[*]{corresponding: xqoy.bahosp@outlook.com}

\begin{abstract}
Atrial fibrillation (AF) is a common cardiac arrhythmia that significantly increases the risk of stroke and heart failure, necessitating reliable and generalizable detection methods from electrocardiogram (ECG) recordings. Although deep learning has advanced automated AF diagnosis, existing approaches often struggle to exploit complementary time–frequency information effectively, limiting both robustness under intra-dataset and generalization across diverse clinical datasets. To address these challenges, we propose a cross-modal deep learning framework comprising two key components: a Bidirectional Gating Module (BGM) and a Cross-modal Supervised Contrastive Learning (CSCL) strategy. The BGM facilitates dynamic, reciprocal refinement between time- and frequency-domain features, enhancing model robustness to signal variations within a dataset. Meanwhile, CSCL explicitly structures the joint embedding space by pulling together label-consistent samples and pushing apart different ones, thereby improving inter-class separability and enabling strong cross-dataset generalization. We evaluate our method using five-fold cross-validation on the AFDB and CPSC2021 datasets. Furthermore, to assess cross-dataset generalization, we conduct bidirectional cross-dataset experiments across AFDB, CPSC2021, LTAF, and SHDBAF by training on one dataset and testing on another. Results show consistent improvements over state-of-the-art methods across multiple metrics, demonstrating that our approach achieves both high intra-dataset robustness and excellent cross-dataset generalization. We further demonstrate that our method achieves high computational efficiency and anti-interference capability, making it suitable for edge deployment. The source code is publicly available at https://github.com/Ou-Young-1999/MGCNet.
\end{abstract}
\begin{document}

\flushbottom
\maketitle
%
%
\thispagestyle{empty}


\section*{Introduction}

Atrial fibrillation (AF) is the most prevalent cardiac arrhythmia worldwide, affecting more than 50 million individuals—a number projected to increase with population aging—and imposes a substantial and growing burden on healthcare systems \cite{intro1}. AF markedly increases the risk of ischemic stroke and is associated with heart failure, cognitive impairment, and increased all-cause mortality \cite{intro2}. Approximately 27\% cases of AF  are asymptomatic or paroxysmal, often escaping detection during routine clinical screening \cite{intro3}. This diagnostic gap highlights the critical need for automated AF detection systems that combine high sensitivity and specificity to enable timely intervention and prevent severe complications.

The electrocardiogram (ECG) provides the main basis for AF diagnosis, offering detailed electrophysiological information \cite{ecg}. In clinical practice, ECG assessments are typically performed using standard 12-lead recordings or ambulatory Holter monitoring. A standard 12-lead ECG captures a short (approximately 10-second) segment of cardiac activity and can confirm AF when characteristic features, such as absent P-waves, irregular f-waves, and an irregular R–R interval, are present \cite{12leadecg}. However, its short duration limits its effectiveness for detecting paroxysmal AF, which occurs intermittently and unpredictably. Holter monitoring addresses this limitation by enabling continuous ECG recording over 24 hours to 7 days through portable devices, making it the preferred method for evaluating suspected paroxysmal AF when standard ECG results are negative \cite{holterecg}. Despite its extended observation window, Holter analysis traditionally depends on manual review, which is labor-intensive, prone to inter-observer variability, and unsuitable for large-scale or real-time applications \cite{ecgai}.

Recent advances in machine learning and deep learning have facilitated automated detection of AF from ECG signals \cite{medai}. However, the majority existing methods rely on features extracted from a single signal domain (either time or frequency) and consequently overlook the complementary electrophysiological characteristics of AF. Clinically, AF manifests through two interrelated patterns: an irregular ventricular response in the time domain and spectrally diffuse atrial activity in the frequency domain \cite{multimodal}. Models confined to one domain capture only a partial view of this arrhythmia, which compromises their robustness and generalizable under real-world recording variability.

A further limitation lies in how multimodal information is integrated and learned. Conventional fusion strategies—such as simple concatenation or fixed-weight averaging—process time- and frequency-domain features independently, ignoring their context-dependent interactions \cite{fusion1}. At the same time, representation learning without explicit supervision often yields embeddings with insufficient class separation, particularly in the presence of noisy. These issues not only limit performance within a given dataset but also hinder model transfer across diverse clinical datasets \cite{fusion2}.





To address these challenges, we propose \textbf{MGCNet (Multimodal Gated Contrastive Network)}, a deep learning architecture for robust and generalizable AF detection from ambulatory ECG recordings. Our approach delivers three key contributions

(1) A \textbf{Bidirectional Gating Module (BGM)} serving as a purpose-built alignment mechanism for cross-domain transfer. Unlike conventional concatenation, BGM uses adaptive bidirectional gating to selectively harmonize modalities, extracting domain-invariant representations stable under varying distributions.

(2) A \textbf{Cross-modal Supervised Contrastive Learning (CSCL)} strategy tightly coupled with BGM. This co-design ensures representation learning and discriminative optimization reinforce each other, overcoming the limitations of generic contrastive methods on heterogeneous ECG data.

(3) System-level validation on four benchmarks (CPSC2021, AFDB, LTAF, SHDBAF). MGCNet consistently outperforms existing methods in both intra- and cross-dataset settings, demonstrating that the non-trivial integration of gated alignment and supervised contrastive learning—rather than any single component—is essential for clinically viable generalization.

\section*{Related Works}

Research on automatic AF detection has undergone a paradigm shift—from rule-based and traditional machine learning approaches to end-to-end deep learning frameworks. Early methods relied heavily on expert-designed features and shallow classifiers, whereas recent advances leverage deep neural networks to learn direct representation from raw or transformed signals \cite{multimodal}.

Traditional AF detection methods relied on handcrafted features, such as RR interval statistics and time–frequency descriptors, to capture AF’s characteristic rhythm irregularity and spectral disorganization, combined with classical classifiers like SVM, random forests, or k-NN \cite{mlecg2}\cite{mlecg3}. While effective on clean, controlled datasets, these approaches were inherently limited by manual feature design and often generalized poorly across diverse devices, patient cohorts, and real-world noise conditions.

The rise of deep learning has shifted AF detection toward end-to-end models that learn directly from ECG signals without relying on handcrafted features. Initial deep learning approaches operated primarily in the time domain. For example, Xia et al. \cite{cnnecg} applied one-dimensional convolutional neural networks (1D-CNNs) to raw ECG waveforms, demonstrating the feasibility of data-driven rhythm classification. Later work incorporated recurrent networks \cite{rnnecg1}\cite{rnnecg2} and Transformer architectures \cite{transecg1}\cite{transecg2} to better model long-range temporal dependencies, while others introduced multi-scale convolutions \cite{msecg1}\cite{msecg2} or attention mechanisms \cite{attenecg} to enhance local and global feature extraction. Li et al. \cite{zshape} took a different route by reshaping 1D ECG segments into a 2D “Z-shaped” layout to take advantage of image-based CNNs.

At the same time, a parallel line of research focused on the frequency domain. These methods first transformed ECG signals into time–frequency representations, such as short-time Fourier transform (STFT) spectrograms or continuous wavelet scalograms, and then applied 2D-CNNs for classification \cite{freqcnn1}\cite{freqcnn2}\cite{freqcnn3}. Others explored nonlinear dynamics–based visualizations: for instance, Zhu et al. \cite{rpecg} converted ECG time series into recurrence plots (RPs), which encode temporal self-similarity, and used a ResNet \cite{resnet} architecture to extract discriminative features for AF detection.

Time- and frequency-domain models have performed well on standard AF detection benchmarks, but each only captures part of the full view: irregular heartbeat timing in the time domain and chaotic atrial activity in the frequency domain. To get a more complete view, recent studies started combining both types of information using multimodal networks that process raw ECG signals together with spectral maps or RR intervals \cite{timefreq1}\cite{timefreq2}.

While recent multimodal methods combine time- and frequency-domain features, they often rely on static fusion and standard classification losses, limiting their robustness to signal variations and generalization across datasets.  Although contrastive learning has shown promise in medical representation learning \cite{contra}\cite{contra2}, none explicitly align temporal and spectral embeddings under a unified supervised objective.  Our MGCNet addresses this by jointly modeling dynamic cross-modal interaction and discriminative feature learning—directly enhancing both robustness and generalization in AF detection.

\section*{Materials and Methods}
\subsection*{Materials}
We utilize four publicly available ECG databases for AF analysis: the MIT-BIH Atrial Fibrillation Database (AFDB), the China Physiological Signal Challenge 2021 (CPSC2021), the Long-Term AF database (LTAF) and the Saitama Heart Database Atrial Fibrillation (SHDBAF). The \textbf{AFDB} \cite{afdb} consists of 25 long-term ECG recordings from patients with documented AF. Among them, 23 records contain two simultaneously recorded ECG signals along with expert-annotated rhythm labels. Records 00735 and 03665 are excluded due to the absence of annotations.  Each recording lasts approximately 10 hours, sampled at 250 Hz. The \textbf{CPSC2021} dataset \cite{cpsc} provides variable-length ECG segments extracted from Lead I and Lead II of long-term Holter recordings, sampled at 200 Hz. The data are released in two phases: Phase I includes 730 recordings from 12 AF patients and 42 non-AF patients; Phase II includes 706 recordings from 37 AF patients and 14 non-AF patients. In total, the dataset comprises 1,436 recordings from 105 unique patients. The \textbf{LTAF} \cite{ltaf} includes 84 long-term ECG recordings of subjects with paroxysmal or sustained AF. Each record contains two simultaneously recorded ECG signals digitized at 128 Hz with 12-bit resolution over a 20 mV range; record durations vary but are typically 24 to 25 hours. The \textbf{SHDBAF} \cite{shdbaf} is a novel open-sourced Holter ECG database from Japan, containing data from 122 unique subjects with paroxysmal atrial fibrillation. Among the 128 recordings, 98 contain raw ECG data with rhythm annotations at the beat level, manually performed by a cardiology fellow. The remaining recordings consist only of ECG traces without annotations.

\begin{figure}[!htbp]
    \centering
    \includegraphics[width=1.0\linewidth]{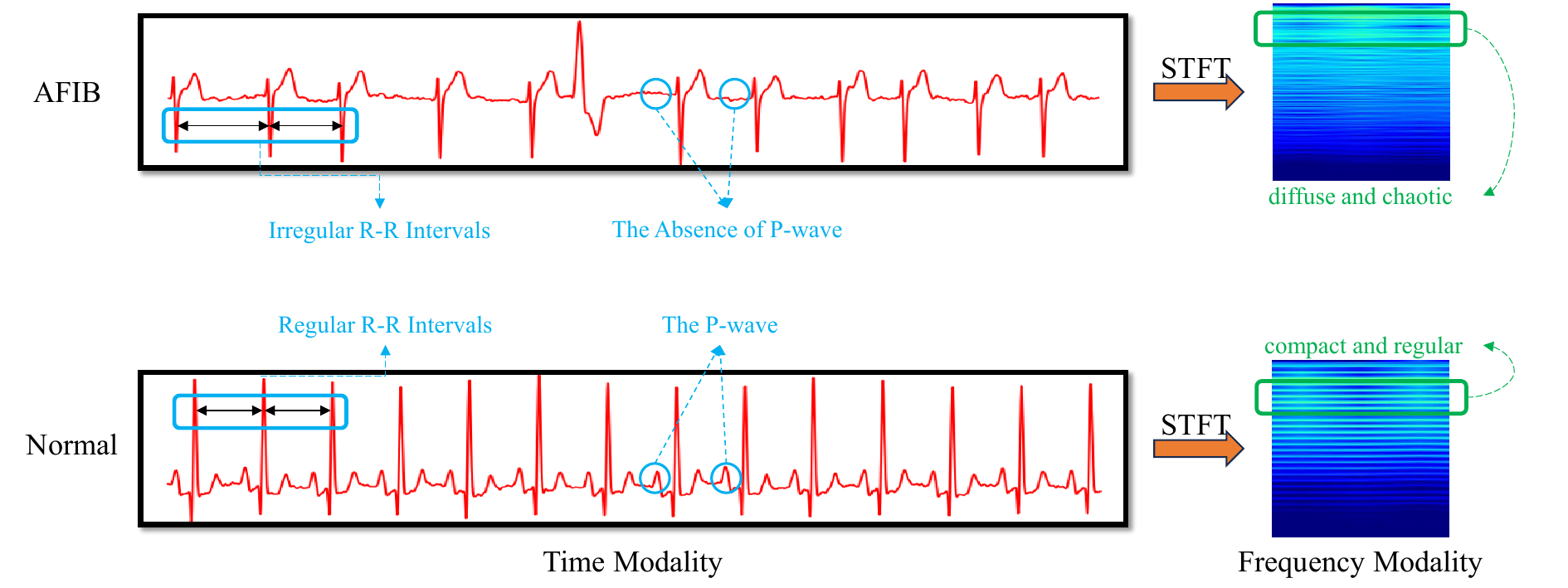}
    \caption{Spectrogram of the cleaned ECG segment generated via STFT, encoded as a three-channel heatmap for frequency-domain modeling. In the time domain, AFIB is characterized by the absence of P waves and highly irregular R-R intervals, and Normal is characterized by the clear P waves and regular R-R intervals. In the frequency domain (STFT spectrogram), AFIB exhibits a diffuse and chaotic spectral pattern, and Normal exhibits a compact and regular spectral pattern.}
    \label{t2f}
\end{figure}

All ECG recordings from the AFDB and the CPSC2021 are uniformly resampled to 250~Hz, and only the first lead is retained. The preprocessing pipeline consists of four sequential steps:

\textbf{Resampling.}  
The input signal ${x} \in \mathbb{R}^L$ is resampled to a fixed sampling frequency $f_s=250$ using linear interpolation. Let $r = f_s / f_{\text{orig}}$ and $L' = \lfloor L r \rfloor$. The resampled signal ${x}' \in \mathbb{R}^{L'}$ is defined on a uniform time grid $t'_i = (L-1) \cdot i / (L'-1)$ for $i = 0, \dots, L'-1$, with values computed as:
\begin{equation}
x'(t'_i) = x(\lfloor t'_i \rfloor) + \bigl(x(\lfloor t'_i \rfloor + 1) - x(\lfloor t'_i \rfloor)\bigr) \cdot (t'_i - \lfloor t'_i \rfloor),
\end{equation}
where $x(t)$ is implicitly zero-padded or clamped at boundaries if needed.

\textbf{Filtering.}  
To suppress baseline wander, powerline interference, and high-frequency noise, each raw ECG signal $ x[n] $ is filtered using a zero-phase fourth-order Butterworth bandpass filter with cutoff frequencies of 0.5~Hz and 40~Hz \cite{filter}. The filtering is implemented in forward--backward mode to eliminate phase distortion:
\begin{equation}
    x_{\text{clean}}[n] = \text{filtfilt}(b_{\text{BP}}, a_{\text{BP}}, x[n]),
\end{equation}
where $ b_{\text{BP}} $ and $ a_{\text{BP}} $ are the numerator and denominator coefficients of the discrete-time bandpass filter, obtained by cascading a low-pass section ($ f_c = 40 $~Hz) and a high-pass section ($ f_c = 0.5 $~Hz). This preserves the morphological integrity of the P-wave, QRS complex, and T-wave while attenuating out-of-band artifacts.

\textbf{Segmentation.}  
The filtered recording is partitioned into non-overlapping 10-second segments \cite{segment}, each containing $ N = 2500 $ samples (since $ 10~\text{s} \times 250~\text{Hz} = 2500 $). Each segment inherits the original label, resulting in a binary classification task between atrial fibrillation (AFIB) and normal sinus rhythm (N).

\textbf{Quality Assessment.}  
To avoid misleading the model with noisy or artifact-contaminated segments, we apply an automated quality assessment based on the method of Zhao et al. \cite{quality1}\cite{quality2}. This approach evaluates both R-peak consistency and ECG waveform morphology to assign a binary label ("qualified" or "unqualified"). Only qualified segments are retained for training and evaluation, ensuring that the model learns from physiologically plausible signals and improving its robustness and generalization in real-world scenarios.

\textbf{Frequency-Domain Modality Generation.}
For each 10-s ECG segment \( x_{\text{clean}}[n] \in \mathbb{R}^{2500} \), we compute a magnitude spectrogram via short-time Fourier transform (STFT) \cite{stft} using a 1-s Hann window (\(N_{\text{win}} = 250\)) and 50\% overlap (\(H = 125\)) at 250~Hz sampling rate. To emphasize physiologically relevant dynamics while suppressing noise, we retain frequencies \(\leq 40\)~Hz—beyond which ECG energy is negligible. The spectrogram \(A[m,k] = |X[m,k]|\) is converted to decibels:
\begin{equation}
    A_{\text{dB}} = 20 \log_{10}(A + \varepsilon), \quad \varepsilon = 10^{-8},
\end{equation}
clipped to \([-80, \max(A_{\text{dB}})]\), and linearly normalized to \([0,1]\). It is then resized to \(128 \times 128\) via anti-aliased bilinear interpolation and mapped to an RGB image using the \texttt{jet} colormap:
\begin{equation}
    \mathbf{I}_{\text{freq}} = \text{Colormap}_{\text{jet}}(\tilde{A}) \in [0,1]^{128 \times 128 \times 3}.
\end{equation}
This representation preserves AF-associated spectral disorganization and serves as input to the spectral branch of our network (Figure.~\ref{t2f}).

The final dataset statistics after preprocessing are summarized in Table~\ref{datasets}.

\begin{table}[htbp]
\centering
\caption{Summary of the AFDB, CPSC2021, LTAF and SHDBAF datasets used in this study.}
\label{datasets}
\begin{tabular}{lcccc}
\toprule
\textbf{Attributes} & \textbf{AFDB} & \textbf{CPSC2021} & \textbf{LTAF}& \textbf{SHDBAF}\\
\hline
Number of patients & 23 & 105 & 84 & 98 \\
Original sampling frequency (Hz) & 250 & 200 & 200 & 200 \\
Resampling frequency (Hz) & 250 & 250 & 250 & 250 \\
Number of segments & 58771 & 137010 & 365714 & 763064\\
Number of 'N' segments & 36718 & 89570 & 187470 & 614413\\
Number of 'AFIB' segments & 22053 & 47440 & 178244 & 148651  \\
\bottomrule
\end{tabular}
\end{table}

\subsection*{Methods}
\subsubsection*{Overall Architecture}
\begin{figure}[!htbp]
    \centering
    \includegraphics[width=1.0\linewidth]{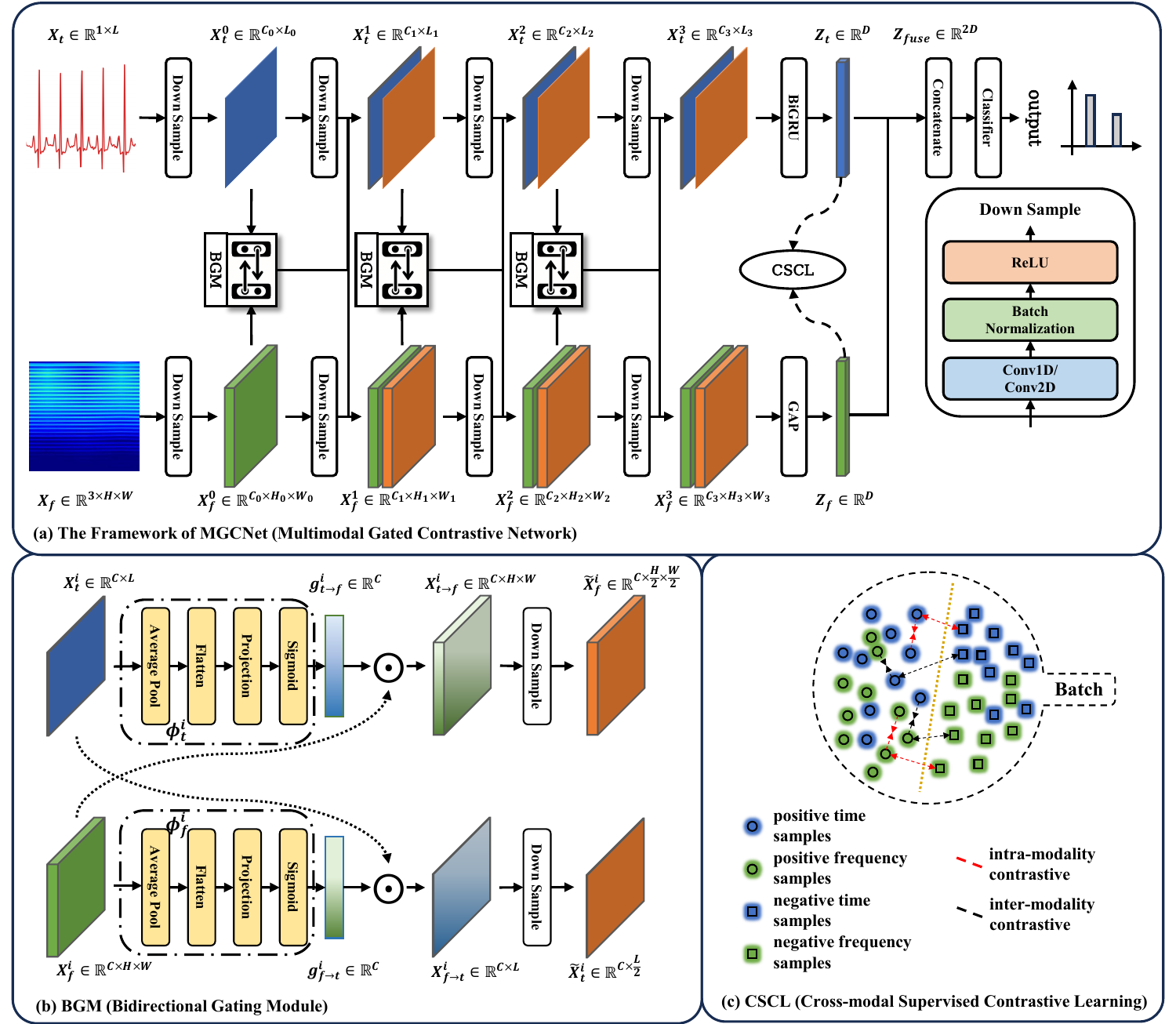}
    \caption{(a) The full multimodal network for AF detection; (b) The BGM enabling dynamic interaction between time- and frequency-domain features; (c) The CSCL that enforces discriminative embedding alignment across modalities.}
    \label{mgcnet}
\end{figure}

Given an input ECG segment $\mathbf{X}_{t} \in \mathbb{R}^{1 \times L}$ of length $L$, our proposed \textbf{MGCNet} processes it through two parallel branches to extract complementary temporal and spectral representations. As illustrated in Figure. \ref{mgcnet}, the framework comprises the following stages:
\textbf{Dual-branch Feature Extraction}: The \textit{time-domain branch} takes the raw ECG signal $\mathbf{X}_{t}$ and extracts a temporal feature map $\mathbf{X}_{t}^{i} \in \mathbb{R}^{C_{i} \times L_{i}}$ using a 1D convolutional neural network (1D-CNN) encoder. The \textit{frequency-domain branch} first converts $\mathbf{X}_{t}$ into a frequency representation via the Short-Time Fourier Transform (STFT) \cite{stft}, resulting in a spectrogram $\mathbf{X}_f \in \mathbb{R}^{3 \times H \times W}$. This spectrogram is then encoded by a 2D-CNN backbone to produce a spectral feature map $\mathbf{X}_{f}^{i} \in \mathbb{R}^{C_{i} \times H_{i} \times W_{i}}$. Through careful design, both branches output features with the same channel dimension $C_{i}$, facilitating subsequent cross-modal interaction. Here, $L_{i}$ denotes the downsampled sequence length in the temporal branch, while $H_{i},W_{i}$ represent the spatial height and width of the spectral feature map after downsampling.

\textbf{Bidirectional Gated Module (BGM)}: 
To facilitate mutual refinement between modalities, the temporal feature map $\mathbf{X}_{t}^{i}$ and the spectral feature map $\mathbf{X}_{f}^{i}$ are jointly processed by a \textit{Bidirectional Gated Module}, which enables adaptive cross-modal information exchange. The module generates enhanced representations $\tilde{\mathbf{X}}_{t}^{i}$ and $\tilde{\mathbf{X}}_{f}^{i}$ that incorporate complementary cues from the other modality. These refined features are then concatenated with their original counterparts along the channel dimension:
\begin{align}
    \mathbf{X}_{t}^{i+1} = [\mathbf{X}_{t}^{i}; \tilde{\mathbf{X}}_{t}^{i}], 
    \mathbf{X}_{f}^{i+1} = [\mathbf{X}_{f}^{i}; \tilde{\mathbf{X}}_{f}^{i}],
\end{align}
where $[\cdot;\cdot]$ denotes channel-wise concatenation, and the stage index $i \in \{0,1,2\}$ corresponds to successive downsampling levels in the encoder.

\textbf{Modality-specific Global Aggregation}:
The augmented temporal feature from the final stage, ${\mathbf{X}}_{t}^{3} \in \mathbb{R}^{C_{3} \times L_{3}}$ is first refined by a channel-mixing 1D convolutional layer to enhance inter-channel interactions, then processed by a bidirectional GRU (BiGRU) \cite{bigru} to capture long-range temporal dynamics. The forward and backward terminal hidden states are concatenated and projected to a D-dimensional space, yielding the final temporal embedding $\mathbf{Z}_t \in \mathbb{R}^{D}$:
\begin{equation}
    \mathbf{Z}_t = \mathrm{BiGRU}({\mathbf{X}}_{t}^{3}).
\end{equation}
Similarly, the augmented spectral feature ${\mathbf{X}}_{f}^{3} \in \mathbb{R}^{C_{3} \times H_{3} \times W_{3}}$ is first refined by a channel-mixing 2D convolutional layer, followed by global average pooling (GAP) over the spatial dimensions to produce the spectral embedding $\mathbf{Z}_f \in \mathbb{R}^{D}$:
\begin{equation}    
    \mathbf{Z}_f = \frac{1}{H_3 \times W_3}\sum_{m=1}^{H_3}\sum_{n=1}^{W_3} \mathbf{X}_{f}^{3}.
\end{equation}

\textbf{Cross-modal Supervised Contrastive Learning (CSCL)}: 
The temporal embedding $\mathbf{Z}_t$ and spectral embedding $\mathbf{Z}_f$ are jointly optimized through a dual-level supervised contrastive loss that operates at both intra-modal and inter-modal levels, guided solely by class labels \cite{supcontra}. The \textit{intra-modal contrastive loss} $\mathcal{L}_{\mathrm{cont}}^{\mathrm{intra}}$ encourages embeddings of the same class within each modality to cluster tightly while separating those of different classes. The \textit{inter-modal contrastive loss} $\mathcal{L}_{\mathrm{cont}}^{\mathrm{inter}}$ treats any temporal–spectral embedding pair sharing the same class label as a positive pair and pulls them closer in the joint embedding space; conversely, pairs with mismatched labels are repelled. The total contrastive objective is defined as:
\begin{equation}
    \mathcal{L}_{\mathrm{cont}} = \frac{1}{2N} \sum_{i=1}^{N} \left( \mathcal{L}_{\mathrm{cont}}^{\mathrm{intra}} + \mathcal{L}_{\mathrm{cont}}^{\mathrm{inter}} \right).
\end{equation}
Here, N represents the number of samples in the current mini-batch.

\textbf{Concatenation and Classification}:
The modality-specific embeddings $\mathbf{Z}_t$ and $\mathbf{Z}_f$ are fused via a concatenation mechanism:
\begin{align}
    \mathbf{Z}_{\mathrm{fuse}} &= [\mathbf{Z}_t; \mathbf{Z}_f].
\end{align}
The fused representation $\mathbf{Z}_{\mathrm{fuse}}$ is passed through a classifier head to produce logits ${p}$, and the classification loss is:
\begin{equation}
    \mathcal{L}_{\mathrm{cls}} = -\frac{1}{N} \sum_{i=1}^{N} \left[ y^{(i)} \log p^{(i)} + (1 - y^{(i)}) \log (1 - p^{(i)}) \right].
\end{equation}
Here, N represents the number of samples in the current mini-batch. $y^{(i)} \in \{0,1\}$ is the true binary label of the i-th sample; $p^{(i)} \in (0,1)$ is the probability that the i-th sample predicted by the model belongs to the positive class. The final loss function is:
\begin{equation}
    \mathcal{L}_{\mathrm{total}} = \mathcal{L}_{\mathrm{cls}} +\lambda \mathcal{L}_{\mathrm{cont}},
\end{equation}
where $\lambda$ is the hyperparameter. The final network structure is shown in Table \ref{architecture}.

\begin{table}[!htbp]
\centering
\caption{Architecture of the proposed MGCNet. $B$ denotes batch size, $N$ denotes number of classifications. All convolutional layers are followed by batch normalization and ReLU activation (not shown for brevity).}
\label{architecture}
\begin{tabular}{lccccc}
\toprule
\textbf{Stage} & \textbf{Module} & \textbf{Time Input} & \textbf{Time Output} & \textbf{Freq Input} & \textbf{Freq Output} \\
\midrule
Initial & Time Init & $[B, 1, 2500]$ & $[B, 32, 1250]$ & — & — \\
        & Freq Init & — & — & $[B, 3, 128, 128]$ & $[B, 32, 64, 64]$ \\
\midrule
Level 1 & BGM-1 & $[B, 32, 1250]$ & $[B, 32, 625]$ & $[B, 32, 64, 64]$ & $[B, 32, 32, 32]$ \\
        & Downsample-1 & $[B, 32, 1250]$ & $[B, 32, 625]$ & $[B, 32, 64, 64]$ & $[B, 32, 32, 32]$ \\
        & Concat + Expand-1 & $[B, 64, 625]$ & $[B, 64, 625]$ & $[B, 64, 32, 32]$ & $[B, 64, 32, 32]$ \\
\midrule
Level 2 & BGM-2 & $[B, 64, 625]$ & $[B, 64, 313]$ & $[B, 64, 32, 32]$ & $[B, 64, 16, 16]$ \\
        & Downsample-2 & $[B, 64, 625]$ & $[B, 64, 313]$ & $[B, 64, 32, 32]$ & $[B, 64, 16, 16]$ \\
        & Concat + Expand-2 & $[B, 128, 313]$ & $[B, 128, 313]$ & $[B, 128, 16, 16]$ & $[B, 128, 16, 16]$ \\
\midrule
Level 3 & BGM-3 & $[B, 128, 313]$ & $[B, 128, 157]$ & $[B, 128, 16, 16]$ & $[B, 128, 8, 8]$ \\
        & Downsample-3 & $[B, 128, 313]$ & $[B, 128, 157]$ & $[B, 128, 16, 16]$ & $[B, 128, 8, 8]$ \\
        & Concat + Expand-3 & $[B, 256, 157]$ & $[B, 256, 157]$ & $[B, 256, 8, 8]$ & $[B, 256, 8, 8]$ \\
\midrule
Fusion & GRU & $[B, 256, 157]$ & $[B, 256]$ & — & — \\
       & Global AvgPool & — & — & $[B, 256, 8, 8]$ & $[B, 256]$ \\
       & Fusion + Classifier & $[B, 512]$ & $[B, N]$ & — & — \\
\bottomrule
\end{tabular}
\end{table}

\subsubsection*{Bidirectional Gated Module}
As shown in Figure.\ref{mgcnet}, the \textbf{BGM} enables adaptive cross-modal interaction while progressively reducing spatio-temporal resolution. At hierarchical stage $i$, given the temporal feature map ${\mathbf{X}}_{t}^{i} \in \mathbb{R}^{C_{i} \times L_{i}}$  and spectral feature map ${\mathbf{X}}_{f}^{i} \in \mathbb{R}^{C_{i} \times H_{i} \times W_{i}}$, BGM generates cross-modal gating signals to selectively enhance informative channels. Specifically, a global context vector is extracted from each modality via adaptive pooling and used to modulate the other.
\begin{align}
    \mathbf{g}_{f \to t}^i &= \phi_f^i(\mathbf{X}_f^i) = \sigma\big( \mathbf{W}_f^i \cdot \mathrm{GAP}_{2D}(\mathbf{X}_f^i) \big),\\
    \mathbf{g}_{t \to f}^i &= \phi_t^i(\mathbf{X}_t^i) = \sigma\big( \mathbf{W}_t^i \cdot \mathrm{GAP}_{1D}(\mathbf{X}_t^i) \big),
\end{align}
where $\mathrm{GAP}_{2D}(\cdot)$ and $\mathrm{GAP}_{1D}(\cdot)$ denote global average pooling over spatial dimensions $(H,W)$ and temporal dimension $L$, respectively; ${\mathbf{W}}_{t}^{i}, {\mathbf{W}}_{f}^{i} \in \mathbb{R}^{C_{i} \times C_{i}}$ are linear projection matrices; and $\sigma$ is the sigmoid activation that produces channel-wise gating weights in $(0,1)$. The gating signals are broadcast across their respective spatial or temporal dimensions and applied multiplicatively to modulate the original features:
\begin{align}
{\mathbf{X}}_{f \to t}^{i} = \mathbf{X}_{t}^{i} \otimes \mathbf{g}_{f \to t}^{i}, 
{\mathbf{X}}_{t \to f}^{i} = \mathbf{X}_{f}^{i} \otimes \mathbf{g}_{t \to f}^{i},
\end{align}
with $\otimes$ indicating broadcast multiplication. Finally, the gated features are downsampled via strided convolutions to produce outputs for the next stage:
\begin{align}
\tilde{\mathbf{X}}_{t}^{i} = \mathrm{Conv1D}\big( \mathbf{X}_{f \to t}^{i} \big), 
\tilde{\mathbf{X}}_{f}^{i} = \mathrm{Conv2D}\big( \mathbf{X}_{t \to f}^{i} \big).
\end{align}

\subsubsection*{Cross-modal Supervised Contrastive Learning}
We adopt the supervised contrastive learning (SupCon) framework \cite{supcontra}, and apply it across modalities: for each labeled ECG segment, its time-domain and frequency-domain embeddings are treated as a positive pair. This cross-domain alignment enforces consistency between complementary signal views, yielding representations that are both robust to modality-specific variations and generalizable across domains, effectively acting as an implicit regularizer.

In order to implement the \textbf{CSCL} module, we employ a dual-level supervised contrastive loss to enforce both intra-modal discrimination and inter-modal alignment. Let $\mathbf{Z}_t^{(i)}$, $\mathbf{Z}_f^{(i)}$ be L2-normalized embeddings of sample $i$ with label $y^{(i)} \in \{0,1\}$, and $\mathrm{sim}(\mathbf{u}, \mathbf{v}) = \frac{\mathbf{u}^\top \mathbf{v}}{\|\mathbf{u}\| \|\mathbf{v}\|}$ denote cosine similarity.
The \textbf{intra-modal loss} pulls together same-class samples within each modality:
\begin{equation}
\mathcal{L}_{\mathrm{cont}}^{\mathrm{intra}} = 
-\log \frac{\sum_{j \neq i,y^{(j)}=y^{(i)}}exp({\mathrm{sim}(\mathbf{z}_t^{(i)}, \mathbf{z}_t^{(j)}) / \tau})}{\sum_{k \neq i} exp({\mathrm{sim}(\mathbf{z}_t^{(i)}, \mathbf{z}_t^{(k)}) / \tau})}
+ (t \leftrightarrow f).
\end{equation}
The \textbf{inter-modal loss} aligns cross-modal representations of the same class:
\begin{equation}
\mathcal{L}_{\mathrm{cont}}^{\mathrm{inter}} = 
-\log \frac{\sum_{j,y^{(j)}=y^{(i)}}exp({\mathrm{sim}(\mathbf{z}_t^{(i)}, \mathbf{z}_f^{(j)}) / \tau})}{\sum_{k=1}^{N} exp({\mathrm{sim}(\mathbf{z}_t^{(i)}, \mathbf{z}_f^{(k)}) / \tau})}
+ (t \leftrightarrow f),
\end{equation}
where $(t \leftrightarrow f)$ denotes the symmetric term for spectral embeddings, temperature $\tau > 0$.

\section*{Results}
\subsection*{Evaluation Metrics}
To comprehensively evaluate the performance of our method on AF detection, we report six standard evaluation metrics: \textit{Accuracy}, \textit{Area Under the ROC Curve (AUC)}, \textit{Precision}, \textit{Recall (Sensitivity)}, \textit{F1-score}, and \textit{Specificity}. Let TP, TN, FP, and FN denote the numbers of true positives, true negatives, false positives, and false negatives, respectively. The metrics are defined as follows:
\begin{itemize}
    \item \textbf{Accuracy} measures the overall fraction of correctly classified samples:$\text{Accuracy} = \frac{\text{TP} + \text{TN}}{\text{TP} + \text{TN} + \text{FP} + \text{FN}}$

    \item \textbf{Precision} (Positive Predictive Value) is the proportion of samples predicted as positive that are truly positive:$\text{Precision} = \frac{\text{TP}}{\text{TP} + \text{FP}}$

    \item \textbf{Recall} (Sensitivity or True Positive Rate) is the proportion of actual positive samples that are correctly identified:$\text{Recall} = \frac{\text{TP}}{\text{TP} + \text{FN}}$

    \item \textbf{Specificity} (True Negative Rate) is the proportion of actual negative samples that are correctly identified:$\text{Specificity} = \frac{\text{TN}}{\text{TN} + \text{FP}}$

    \item \textbf{F1-score} is the harmonic mean of Precision and Recall, providing a balanced measure of the two:$\text{F1} = 2 \cdot \frac{\text{Precision} \cdot \text{Recall}}{\text{Precision} + \text{Recall}}$

    \item \textbf{AUC} (Area Under the ROC Curve) quantifies the model’s ability to discriminate between positive and negative classes across all decision thresholds. It is computed as the area under the curve of True Positive Rate (TPR = Recall) versus False Positive Rate (FPR = $1 - \text{Specificity}$). An AUC value closer to 1.0 indicates better discriminative performance.
\end{itemize}

\begin{figure}[!htbp]
    \centering
    \includegraphics[width=1.0\linewidth]{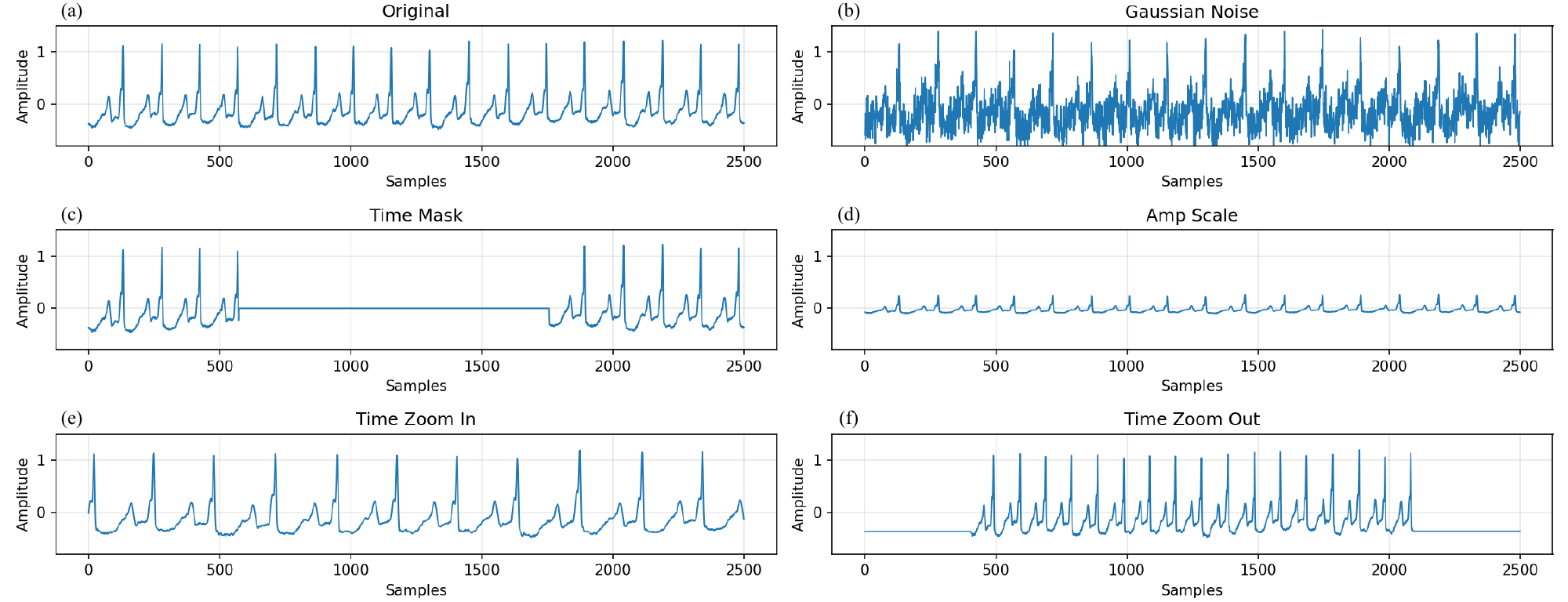}
    \caption{Illustration of the ECG data augmentation \cite{augment} strategies applied in our experiments. 
(a) Original 10-second ECG segment; 
(b) with added Gaussian noise; 
(c) with random time masking; 
(d) with amplitude scaling; 
(e) with time zoom in; 
(f) with time zoom out.}
    \label{augment}
\end{figure}

\begin{table}[!htbp]
\centering
\caption{Performance comparison of different models on AFDB and CPSC2021. The best results in each setting are highlighted in bold.}
\label{comparison}
\begin{tabular}{lcccccc}
\hline
\textbf{Model} & \multicolumn{6}{c}{\textbf{Evaluation Metrics | Mean (standard deviation)}} \\
\cline{2-7}
& \textbf{Acc} & \textbf{AUC} & \textbf{F1} & \textbf{Precision} & \textbf{Recall} & \textbf{Specificity} \\
\hline
\multicolumn{7}{c}{\textbf{Dataset: AFDB}} \\
\hline
SCCNN(2023)\cite{zshape} &
\begin{tabular}[c]{@{}c@{}}0.9614 \\ (0.0198)\end{tabular} &
\begin{tabular}[c]{@{}c@{}}0.9907 \\ (0.0050)\end{tabular} &
\begin{tabular}[c]{@{}c@{}}0.9481 \\ (0.0279)\end{tabular} &
\begin{tabular}[c]{@{}c@{}}0.9473 \\ (0.0285)\end{tabular} &
\begin{tabular}[c]{@{}c@{}}0.9502 \\ (0.0427)\end{tabular} &
\begin{tabular}[c]{@{}c@{}}0.9681 \\ (0.0172)\end{tabular} \\

IMCResNet(2024)\cite{msecg1} &
\begin{tabular}[c]{@{}c@{}}0.9514 \\ (0.0362)\end{tabular} &
\begin{tabular}[c]{@{}c@{}}0.9836 \\ (0.0175)\end{tabular} &
\begin{tabular}[c]{@{}c@{}}0.9356 \\ (0.0472)\end{tabular} &
\begin{tabular}[c]{@{}c@{}}0.9412 \\ (0.0503)\end{tabular} &
\begin{tabular}[c]{@{}c@{}}0.9318 \\ (0.0581)\end{tabular} &
\begin{tabular}[c]{@{}c@{}}0.9631 \\ (0.0350)\end{tabular} \\

MoETransformer(2024)\cite{transecg1} &
\begin{tabular}[c]{@{}c@{}}0.9453 \\ (0.0309)\end{tabular} &
\begin{tabular}[c]{@{}c@{}}0.9857 \\ (0.0096)\end{tabular} &
\begin{tabular}[c]{@{}c@{}}0.9259 \\ (0.0431)\end{tabular} &
\begin{tabular}[c]{@{}c@{}}0.9363 \\ (0.0450)\end{tabular} &
\begin{tabular}[c]{@{}c@{}}0.9170 \\ (0.0543)\end{tabular} &
\begin{tabular}[c]{@{}c@{}}0.9622 \\ (0.0265)\end{tabular} \\

SeqAFNet(2024)\cite{transecg2} &
\begin{tabular}[c]{@{}c@{}}0.9374 \\ (0.0341)\end{tabular} &
\begin{tabular}[c]{@{}c@{}}0.9822 \\ (0.0127)\end{tabular} &
\begin{tabular}[c]{@{}c@{}}0.9170 \\ (0.0462)\end{tabular} &
\begin{tabular}[c]{@{}c@{}}0.9197 \\ (0.0619)\end{tabular} &
\begin{tabular}[c]{@{}c@{}}0.9212 \\ (0.0780)\end{tabular} &
\begin{tabular}[c]{@{}c@{}}0.9473 \\ (0.0450)\end{tabular} \\

MFEGNet(2025)\cite{attenecg} &
\begin{tabular}[c]{@{}c@{}}0.9794 \\ (0.0132)\end{tabular} &
\begin{tabular}[c]{@{}c@{}}0.9957 \\ (0.0045)\end{tabular} &
\begin{tabular}[c]{@{}c@{}}0.9730 \\ (0.0169)\end{tabular} &
\begin{tabular}[c]{@{}c@{}}0.9697 \\ (0.0247)\end{tabular} &
\begin{tabular}[c]{@{}c@{}}0.9767 \\ (0.0209)\end{tabular} &
\begin{tabular}[c]{@{}c@{}}0.9809 \\ (0.0156)\end{tabular} \\

MSCGN(2026)\cite{msecg2} &
\begin{tabular}[c]{@{}c@{}}0.9812 \\ (0.0155)\end{tabular} &
\begin{tabular}[c]{@{}c@{}}0.9965 \\ (0.0042)\end{tabular} &
\begin{tabular}[c]{@{}c@{}}0.9755 \\ (0.0193)\end{tabular} &
\begin{tabular}[c]{@{}c@{}}0.9771 \\ (0.0311)\end{tabular} &
\begin{tabular}[c]{@{}c@{}}0.9774 \\ (0.0164)\end{tabular} &
\begin{tabular}[c]{@{}c@{}}0.9832 \\ (0.0209)\end{tabular} \\

\textbf{MGCNet (Ours)} &
\begin{tabular}[c]{@{}c@{}}\textbf{0.9878} \\ \textbf{(0.0094)}\end{tabular} &
\begin{tabular}[c]{@{}c@{}}\textbf{0.9987} \\ \textbf{(0.0013)}\end{tabular} &
\begin{tabular}[c]{@{}c@{}}\textbf{0.9841} \\ \textbf{(0.0120)}\end{tabular} &
\begin{tabular}[c]{@{}c@{}}\textbf{0.9856} \\ \textbf{(0.0219)}\end{tabular} &
\begin{tabular}[c]{@{}c@{}}\textbf{0.9830} \\ \textbf{(0.0170)}\end{tabular} &
\begin{tabular}[c]{@{}c@{}}\textbf{0.9906} \\ \textbf{(0.0147)}\end{tabular} \\
\hline
\multicolumn{7}{c}{\textbf{Dataset: CPSC2021}} \\
\hline
SCCNN(2023)\cite{zshape} &
\begin{tabular}[c]{@{}c@{}}0.9713 \\ (0.0087)\end{tabular} &
\begin{tabular}[c]{@{}c@{}}0.9900 \\ (0.0136)\end{tabular} &
\begin{tabular}[c]{@{}c@{}}0.9588 \\ (0.0119)\end{tabular} &
\begin{tabular}[c]{@{}c@{}}0.9576 \\ (0.0239)\end{tabular} &
\begin{tabular}[c]{@{}c@{}}0.9605 \\ (0.0136)\end{tabular} &
\begin{tabular}[c]{@{}c@{}}0.9767 \\ (0.0148)\end{tabular} \\

IMCResNet(2024)\cite{msecg1} &
\begin{tabular}[c]{@{}c@{}}0.9337 \\ (0.0204)\end{tabular} &
\begin{tabular}[c]{@{}c@{}}0.9819 \\ (0.0113)\end{tabular} &
\begin{tabular}[c]{@{}c@{}}0.9253 \\ (0.0299)\end{tabular} &
\begin{tabular}[c]{@{}c@{}}0.9024 \\ (0.0444)\end{tabular} &
\begin{tabular}[c]{@{}c@{}}0.9096 \\ (0.0273)\end{tabular} &
\begin{tabular}[c]{@{}c@{}}0.9459 \\ (0.0293)\end{tabular} \\

MoETransformer(2024)\cite{transecg1} &
\begin{tabular}[c]{@{}c@{}}0.9524 \\ (0.0183)\end{tabular} &
\begin{tabular}[c]{@{}c@{}}0.9908 \\ (0.0062)\end{tabular} &
\begin{tabular}[c]{@{}c@{}}0.9307 \\ (0.0291)\end{tabular} &
\begin{tabular}[c]{@{}c@{}}0.9287 \\ (0.0268)\end{tabular} &
\begin{tabular}[c]{@{}c@{}}0.9338 \\ (0.0435)\end{tabular} &
\begin{tabular}[c]{@{}c@{}}0.9622 \\ (0.0139)\end{tabular} \\

SeqAFNet(2024)\cite{transecg2} &
\begin{tabular}[c]{@{}c@{}}0.9539 \\ (0.0181)\end{tabular} &
\begin{tabular}[c]{@{}c@{}}0.9914 \\ (0.0064)\end{tabular} &
\begin{tabular}[c]{@{}c@{}}0.9341 \\ (0.0239)\end{tabular} &
\begin{tabular}[c]{@{}c@{}}0.9352 \\ (0.0524)\end{tabular} &
\begin{tabular}[c]{@{}c@{}}0.9369 \\ (0.0418)\end{tabular} &
\begin{tabular}[c]{@{}c@{}}0.9623 \\ (0.0361)\end{tabular} \\

MFEGNet(2025)\cite{attenecg} &
\begin{tabular}[c]{@{}c@{}}0.9797 \\ (0.0177)\end{tabular} &
\begin{tabular}[c]{@{}c@{}}0.9972 \\ (0.0042)\end{tabular} &
\begin{tabular}[c]{@{}c@{}}\textbf{0.9719} \\ \textbf{(0.0226)}\end{tabular} &
\begin{tabular}[c]{@{}c@{}}0.9610 \\ (0.0460)\end{tabular} &
\begin{tabular}[c]{@{}c@{}}\textbf{0.9846} \\ \textbf{(0.0113)}\end{tabular} &
\begin{tabular}[c]{@{}c@{}}0.9767 \\ (0.0297)\end{tabular} \\

MSCGN(2026)\cite{msecg2} &
\begin{tabular}[c]{@{}c@{}}0.9776 \\ (0.0140)\end{tabular} &
\begin{tabular}[c]{@{}c@{}}0.9973 \\ (0.0032)\end{tabular} &
\begin{tabular}[c]{@{}c@{}}0.9684 \\ (0.0184)\end{tabular} &
\begin{tabular}[c]{@{}c@{}}0.9611 \\ (0.0357)\end{tabular} &
\begin{tabular}[c]{@{}c@{}}0.9766 \\ (0.0072)\end{tabular} &
\begin{tabular}[c]{@{}c@{}}0.9778 \\ (0.0222)\end{tabular} \\

\textbf{MGCNet (Ours)} &
\begin{tabular}[c]{@{}c@{}}\textbf{0.9801} \\ \textbf{(0.0042)}\end{tabular} &
\begin{tabular}[c]{@{}c@{}}\textbf{0.9979} \\ \textbf{(0.0010)}\end{tabular} &
\begin{tabular}[c]{@{}c@{}}0.9715 \\ (0.0067)\end{tabular} &
\begin{tabular}[c]{@{}c@{}}\textbf{0.9639} \\ \textbf{(0.0058)}\end{tabular} &
\begin{tabular}[c]{@{}c@{}}0.9794 \\ (0.0124)\end{tabular} &
\begin{tabular}[c]{@{}c@{}}\textbf{0.9806} \\ \textbf{(0.0023)}\end{tabular} \\
\hline
\end{tabular}
\end{table}

\subsection*{Implementation Details}
All experiments were implemented in PyTorch 2.4.1 with CUDA 11.5 acceleration and conducted on a workstation equipped with an NVIDIA GeForce RTX 3090 GPU (24 GB VRAM) and an Intel Core i7-12700KF CPU. To enhance all model robustness against real-world signal variability, we applied four data augmentation strategies \cite{augment} to the training set (see Figure. \ref{augment} for visual examples): 
(i) additive Gaussian noise with standard deviation $\sigma \in [0, 0.2]$;  
(ii) random masking with a masking ratio uniformly sampled from $[0, 0.1]$;  
(iii) amplitude scaling by a factor drawn from $\mathcal{U}(0.5, 1.5)$; and  
(iv) time-axis zooming (stretching or compression) with scale factors in $[0.8, 1.2]$, implemented via linear interpolation.

The model was trained for 25 epochs with a batch size of 128. We used the Adam optimizer \cite{adam} with an initial learning rate of $1 \times 10^{-3}$, weight decay of $1 \times 10^{-4}$, and an inverse-time learning rate scheduler defined as:$\eta_t = \eta_0/({1 + \gamma \cdot t}),$ where $\eta_0 = 10^{-3}$, decay rate $\gamma = 10^{-4}$, and $t$ denotes the epoch index. The contrastive loss hyperparameters were configured as $\lambda=0.01$ and $\tau=0.1$.

Critically, we adopted a patient-wise five-fold cross-validation protocol: all segments derived from the same patient were assigned exclusively to one fold.    This strategy eliminated inter-patient data leakage and ensures that performance metrics reflect robustness to unseen individuals. Furthermore, patient assignment was stratified to maintain balanced atrial fibrillation (AF) and normal sinus rhythm (N) proportions across all folds.

To further assess the model’s generalization to domain shift, we performed bidirectional cross-dataset evaluation.  Specifically, we trained the model on one dataset and directly evaluated it on another without any fine-tuning or adaptation.  Six scenarios were considered: (i) training on the AFDB and testing on CPSC2021; (ii) training on the AFDB and testing on LTAF;(iii) training on the AFDB and testing on SHDBAF; (ix) training on CPSC2021 and testing on AFDB; (x) training on CPSC2021 and testing on LTAF; (xi) training on CPSC2021 and testing on SHDBAF.

\begin{figure}[!htbp]
    \centering
    \includegraphics[width=1.0\linewidth]{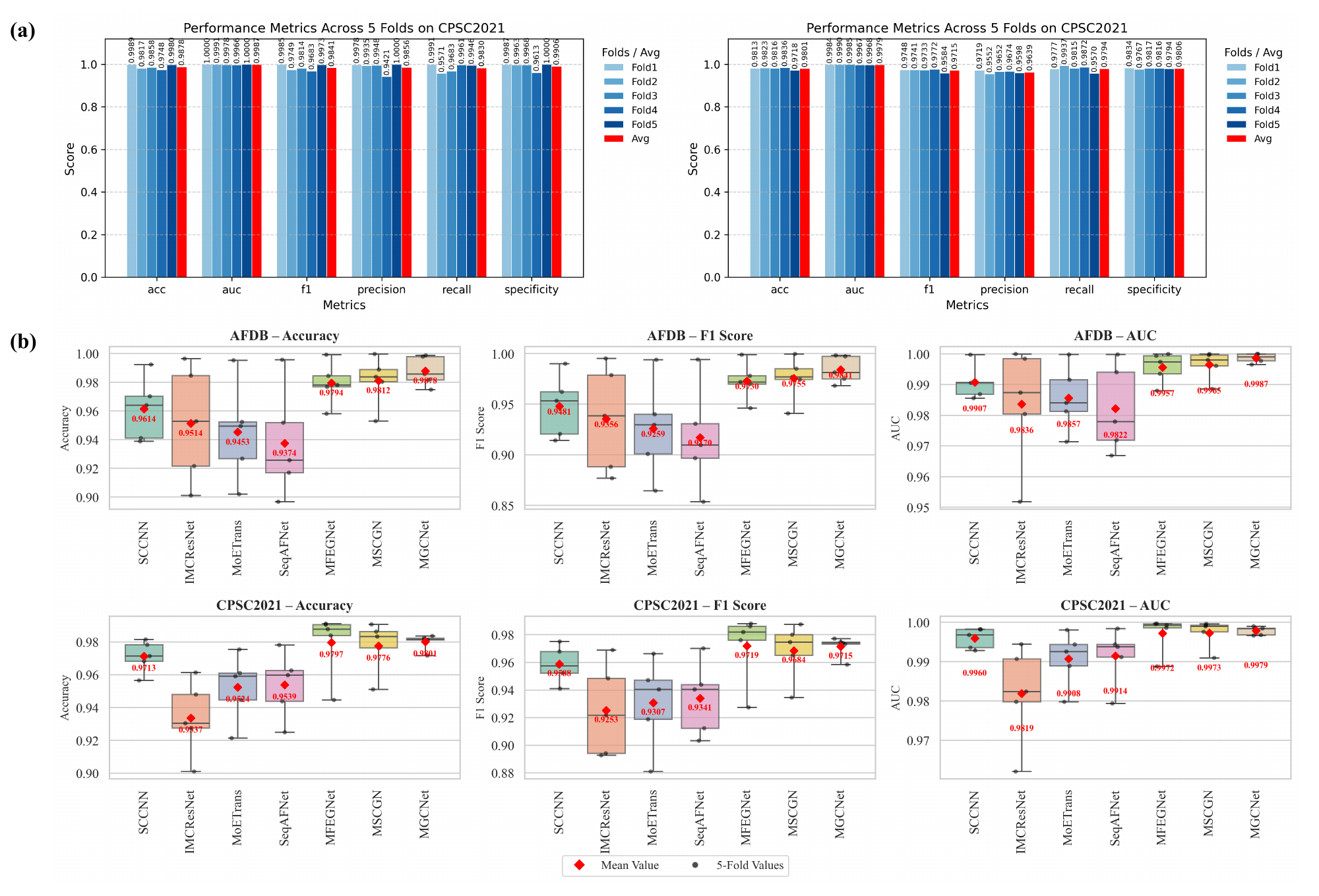}
    \caption{Performance evaluation of MGCNet and comparison with baseline methods.
(a) Five-fold cross-validation results of MGCNet on the AFDB and CPSC2021 across six metrics (Accuracy, AUC, F1-score, Precision, Recall, Specificity).
(b) Box plots comparing the Accuracy, F1-score, and AUC of MGCNet against other state-of-the-art methods.}
    \label{5fold}
\end{figure}

\subsection*{Comparative Experiments}
To validate the effectiveness of the proposed \textbf{MGCNet}, we conducted comprehensive experiments on two widely used ECG datasets: AFDB and CPSC2021. Following standard practice, we performed patient-wise five-fold cross-validation to prevent data leakage across subjects. For a fair comparison, all competing methods were trained and evaluated under identical experimental settings, including the same data preprocessing pipeline and data augmentation strategies. We compared our method against six representative baseline approaches:  

(1) \textbf{SCCNN} \cite{zshape}, which reshapes 1D ECG signals into a 2D "Z-shaped" representation and processes them with a 2D CNN;  

(2) \textbf{IMCResNet} \cite{msecg1}, a multi-scale convolutional architecture that captures interactions across different receptive fields;  

(3) \textbf{MoETransformer} \cite{transecg1}, which integrates Mixture-of-Experts (MoE) modules into a standard Transformer to enhance representational diversity;  

(4) \textbf{SeqAFNet} \cite{transecg2}, a dual-path RNN framework designed to jointly model local and global temporal dependencies;  

(5) \textbf{MFEGNet} \cite{attenecg}, which employs multi-scale gated attention mechanisms for adaptive feature refinement;  

and (6) \textbf{MSCGN} \cite{msecg2}, a network incorporating complementary multi-scale contextual modules.

As shown in Table \ref{comparison} , our method achieved the best performance on four out of six evaluation metrics on both AFDB and CPSC2021, demonstrating its consistent superiority across diverse assessment criteria. Notably, MGCNet attained the highest accuracy of \textbf{0.9878} on AFDB and \textbf{0.9801} on CPSC2021 among all compared methods, highlighting its strong robustness in real-world arrhythmia detection scenarios.

To evaluate the stability and reproducibility of MGCNet, we presented its five-fold cross-validation results on the AFDB and CPSC2021 datasets in Figure~\ref{5fold}. Specifically, Figure~\ref{5fold}(a) displayed bar plots of the average performance across six metrics—Accuracy, AUC, F1-score, Precision, Recall, and Specificity—demonstrating consistently high scores on both datasets. Meanwhile, Figure~\ref{5fold}(b) provided box plots comparing MGCNet against baseline methods in terms of Accuracy, F1-score, and AUC, highlighting its superior and more concentrated performance distribution. As further confirmed by the standard deviations reported in Table~\ref{comparison} (all around 0.02 across metrics and datasets), MGCNet exhibited strong robustness and reliability under patient-wise data partitioning.

\subsection*{Generalization Experiments}
To evaluate generalization under diverse distributional shifts, we conducted cross-dataset experiments across six transfer scenarios: AFDB$\to$CPSC2021, AFDB$\to$LTAF, AFDB$\to$SHDBAF, CPSC2021$\to$AFDB, CPSC2021$\to$LTAF, and CPSC2021$\to$SHDBAF. Here, the left side of each arrow denoted the training dataset and the right side denoted the testing dataset. These settings encompassed substantial heterogeneity in sampling rates and patient demographics, providing a comprehensive benchmark for domain generalization in AF screening.

Table~\ref{generalization} presented the average performance across all six transfers. MGCNet outperformed all competing methods on five of six metrics, achieving an average accuracy of \textbf{0.9296}, AUC of \textbf{0.9691}, precision of \textbf{0.9045}, specificity of \textbf{0.8877}, and F1-score of \textbf{0.9357}. Compared with the second-best MSCGN~\cite{msecg2}, MGCNet improved average accuracy by 3.62\% and F1-score by 3.74\%. Although MFEGNet~\cite{attenecg} attained marginally higher recall, it suffered from substantially lower precision and specificity, yielding a lower F1-score. 

To ensure robust generalization across domains, the model adopted a more conservative decision boundary to mitigate false positives induced by distribution shifts, which inevitably led to the exclusion of some ambiguous positive samples and a consequent decrease in recall. However, this trade-off yield superior precision and specificity, which were clinically critical in wearable or resource-constrained monitoring to prevent alert fatigue and unnecessary referrals. The consistent advantage in these metrics suggests that the BGM and CSCL modules effectively extracted domain-invariant, physiologically grounded features rather than dataset-specific artifacts. Collectively, these results demonstrated that MGCNet possesses superior generalization and clinically appropriate robustness—a prerequisite for multi-center AF screening deployment.

\begin{table*}[!htbp]
\centering
\caption{Cross-dataset generalization performance of MGCNet and competing methods across six distinct scenarios, including the average results. Results are reported for six different transfer settings. The best result in each setting and metric is highlighted in \textbf{bold}.}
\label{generalization}
\begin{tabular}{l|l|cccccc}
\hline
\textbf{Setting} & \textbf{Model} & \textbf{Acc} & \textbf{AUC} & \textbf{F1} & \textbf{Precision} & \textbf{Recall} & \textbf{Specificity} \\
\hline
{\textbf{AFDB} $\to$ \textbf{CPSC2021}} 
& SCCNN (2023)\cite{zshape} & 0.8386 & 0.9497 & 0.8021 & 0.6969 & 0.9448 & 0.7823 \\
& IMCResNet (2024)\cite{msecg1} & 0.8504 & 0.9319 & 0.7982 & 0.7489 & 0.8545 & 0.8482 \\
& MoETransformer (2024)\cite{transecg1} & 0.8681 & 0.9355 & 0.8165 & 0.7878 & 0.8474 & 0.8971 \\
& SeqAFNet (2024)\cite{transecg2} & 0.8646 & 0.9382 & 0.8165 & 0.7691 & 0.8701 & 0.8617 \\
& MFEGNet (2025)\cite{attenecg} & 0.8843 & 0.9593 & 0.8515 & 0.7663 & \textbf{0.9579} & 0.8453 \\
& MSCGN (2026)\cite{msecg2} & 0.8951 & 0.9609 & 0.8611 & 0.7947 & 0.9397 & 0.8714 \\
& \textbf{MGCNet (Ours)} & \textbf{0.9165} & \textbf{0.9643} & \textbf{0.8819} & \textbf{0.8639} & 0.9007 & \textbf{0.9248} \\
\hline
{\textbf{AFDB} $\to$ \textbf{LTAF}} 
& SCCNN (2023)\cite{zshape} & 0.8763 & 0.9679 & 0.8844 & 0.8120 & 0.9710 & 0.7862 \\
& IMCResNet (2024)\cite{msecg1} & 0.9156 & 0.9722 & 0.9180 & 0.8715 & 0.9697 & 0.8641 \\
& MoETransformer (2024)\cite{transecg1} & 0.8832 & 0.9512 & 0.8760 & 0.9076 & 0.8466 & 0.9181 \\
& SeqAFNet (2024)\cite{transecg2} & 0.9242 & 0.9717 & 0.9261 & 0.8826 & 0.9741 & 0.8768 \\
& MFEGNet (2025)\cite{attenecg} & 0.9299 & 0.9781 & 0.9305 & 0.9010 & 0.9619 & 0.8995 \\
& MSCGN (2026)\cite{msecg2} & 0.9484 & \textbf{0.9846} & 0.9485 & 0.9223 & \textbf{0.9763} & 0.9218 \\
& \textbf{MGCNet (Ours)} & \textbf{0.9491} & 0.9778 & \textbf{0.9488} & \textbf{0.9304} & 0.9681 & \textbf{0.9496} \\
\hline
{\textbf{AFDB} $\to$ \textbf{SHDBAF}} 
& SCCNN (2023)\cite{zshape} & 0.7799 & 0.8681 & 0.6954 & 0.6037 & 0.8199 & 0.7911 \\
& IMCResNet (2024)\cite{msecg1} & 0.7777 & 0.8571 & 0.6824 & 0.6068 & 0.7795 & 0.7782 \\
& MoETransformer (2024)\cite{transecg1} & 0.7397 & 0.8774 & 0.6788 & 0.5457 & 0.8976 & 0.7838 \\
& SeqAFNet (2024)\cite{transecg2} & 0.8290 & 0.9046 & 0.7661 & 0.6593 & 0.9142 & 0.8528 \\
& MFEGNet (2025)\cite{attenecg} & 0.8437 & 0.9269 & 0.7901 & 0.6711 & 0.9605 & 0.8763 \\
& MSCGN (2026)\cite{msecg2} & 0.7733 & 0.8792 & 0.7189 & 0.5796 & 0.9463 & 0.8216 \\
& \textbf{MGCNet (Ours)} & \textbf{0.8930} & \textbf{0.9444} & \textbf{0.8479} & \textbf{0.7513} & \textbf{0.9728} & \textbf{0.9153} \\
\hline
{\textbf{CPSC2021} $\to$ \textbf{AFDB}} 
& SCCNN (2023)\cite{zshape} & 0.8413 & 0.9140 & 0.7907 & 0.7828 & 0.7987 & 0.8669 \\
& IMCResNet (2024)\cite{msecg1} & 0.7706 & 0.8240 & 0.7274 & 0.6564 & 0.8157 & 0.7435 \\
& MoETransformer (2024)\cite{transecg1} & 0.7490 & 0.8535 & 0.6840 & 0.6483 & 0.7238 & 0.7642 \\
& SeqAFNet (2024)\cite{transecg2} & 0.8297 & 0.9272 & 0.7870 & 0.6940 & 0.9088 & 0.7878 \\
& MFEGNet (2025)\cite{attenecg} & 0.8756 & 0.9620 & 0.8450 & 0.7935 & 0.9037 & 0.8588 \\
& MSCGN (2026)\cite{msecg2} & 0.9164 & 0.9504 & 0.8947 & 0.8483 & \textbf{0.9465} & 0.8983 \\
& \textbf{MGCNet (Ours)} & \textbf{0.9507} & \textbf{0.9894} & \textbf{0.9331} & \textbf{0.9514} & 0.9154 & \textbf{0.9719} \\
\hline
{\textbf{CPSC2021} $\to$ \textbf{LTAF}} 
& SCCNN (2023)\cite{zshape} & 0.9359 & \textbf{0.9810} & 0.9312 & 0.9766 & 0.8899 & 0.9348 \\
& IMCResNet (2024)\cite{msecg1} & 0.8965 & 0.9574 & 0.8955 & 0.8813 & 0.9102 & 0.8968 \\
& MoETransformer (2024)\cite{transecg1} & 0.9227 & 0.9731 & 0.9173 & 0.9584 & 0.8796 & 0.9217 \\
& SeqAFNet (2024)\cite{transecg2} & 0.9302 & 0.9741 & 0.9254 & 0.9653 & 0.8887 & 0.9292 \\
& MFEGNet (2025)\cite{attenecg} & 0.9380 & 0.9753 & 0.9364 & 0.9489 & 0.9242 & 0.9377 \\
& MSCGN (2026)\cite{msecg2} & 0.9162 & 0.9773 & 0.9112 & 0.9596 & 0.8675 & 0.9350 \\
& \textbf{MGCNet (Ours)} & \textbf{0.9476} & 0.9764 & \textbf{0.9456} & \textbf{0.9560} & \textbf{0.9354} & \textbf{0.9473} \\
\hline
{\textbf{CPSC2021} $\to$ \textbf{SHDBAF}} 
& SCCNN (2023)\cite{zshape} & 0.8632 & 0.9390 & 0.7992 & 0.7260 & 0.8888 & 0.8703 \\
& IMCResNet (2024)\cite{msecg1} & 0.8662 & 0.9431 & 0.8026 & 0.7323 & 0.8878 & 0.8722 \\
& MoETransformer (2024)\cite{transecg1} & 0.8651 & 0.9378 & 0.8026 & 0.7276 & \textbf{0.8948} & 0.8734 \\
& SeqAFNet (2024)\cite{transecg2} & 0.8691 & 0.9313 & 0.8036 & 0.7438 & 0.8738 & 0.8704 \\
& MFEGNet (2025)\cite{attenecg} & 0.8965 & 0.9624 & 0.8521 & 0.8020 & 0.9088 & 0.9000 \\
& MSCGN (2026)\cite{msecg2} & 0.9107 & 0.9181 & 0.8679 & 0.8606 & 0.8754 & 0.9009 \\
& \textbf{MGCNet (Ours)} & \textbf{0.9204} & \textbf{0.9625} & \textbf{0.8696} & \textbf{0.8732} & 0.8660 & \textbf{0.9053} \\
\hline
{\textbf{Average}}
& SCCNN (2023)\cite{zshape} & 0.8559 & 0.9366 & 0.8172 & 0.7663 & 0.8855 & 0.8386 \\
& IMCResNet (2024)\cite{msecg1} & 0.8462 & 0.9143 & 0.8040 & 0.7495 & 0.8696 & 0.8338 \\
& MoETransformer (2024)\cite{transecg1} & 0.8380 & 0.9214 & 0.7959 & 0.7626 & 0.8483 & 0.8597 \\
& SeqAFNet (2024)\cite{transecg2} & 0.8745 & 0.9412 & 0.8375 & 0.7857 & 0.9050 & 0.8631 \\
& MFEGNet (2025)\cite{attenecg} & 0.8947 & 0.9607 & 0.8676 & 0.8138 & \textbf{0.9362} & 0.8863 \\
& MSCGN (2026)\cite{msecg2} & 0.8934 & 0.9451 & 0.8671 & 0.8275 & 0.9253 & 0.8915 \\
& \textbf{MGCNet (Ours)} & \textbf{0.9296} & \textbf{0.9691} & \textbf{0.9045} & \textbf{0.8877} & 0.9264 & \textbf{0.9357} \\
\hline
\end{tabular}
\end{table*}

\subsection*{Ablation Experiments}
We further conducted an ablation study to evaluate the impact of key components, namely the BGM and CSCL, on both intra-dataset performance and cross-dataset generalization. All variants were trained under identical protocols.

As shown in Table \ref{ablation}, the full MGCNet consistently achieved the highest accuracy and AUC across all settings. Removing BGM led to noticeable performance degradation, particularly under domain shift, where both accuracy and AUC dropped substantially compared to the full model. This highlighted BGM’s role in aligning time-frequency representations across heterogeneous data distributions. Similarly, ablating CSCL resulted in reduced discriminability, with pronounced declines observed in cross-dataset scenarios than in intra-dataset evaluation. This suggested that CSCL not only improves feature quality within a dataset but also enhances robustness to distributional shifts by promoting consistent hierarchical representations.

Single-branch variants (time-only or frequency-only) performed significantly worse, especially in cross-dataset tests, confirming that multimodal fusion is essential for reliable AF detection across diverse clinical recording conditions.

Overall, both BGM and CSCL contributed critically to strong robustness and generalization, as evidenced by consistent gains in both accuracy and AUC, which demonstrating their value for real-world ECG monitoring applications.

\begin{table}[htbp]
\centering
\caption{Ablation study of key components in MGCNet under intra-dataset and cross-dataset settings. Results are reported as accuracy / AUC.}
\label{ablation}
\begin{tabular}{l|c|c|c|c}
\hline
{\textbf{Model Variant}} 
& \multicolumn{2}{c|}{\textbf{Intra-Dataset}} 
& \multicolumn{2}{c}{\textbf{Cross-Dataset}} \\
\cline{2-5}
& AFDB & CPSC2021 & AFDB $\to$ CPSC2021 & CPSC2021 $\to$ AFDB \\
\hline
Full MGCNet (Ours)          & \textbf{0.9878 / 0.9987} & \textbf{0.9801 / 0.9979} & \textbf{0.9165 / 0.9643} & \textbf{0.9507 / 0.9894} \\
w/o BGM                     & 0.9732 / 0.9961 & 0.9645 / 0.9952 & 0.8721 / 0.9487 & 0.8934 / 0.9621 \\
w/o CSCL                    & 0.9785 / 0.9970 & 0.9710 / 0.9963 & 0.8893 / 0.9532 & 0.9102 / 0.9745 \\
Time-only                   & 0.9621 / 0.9934 & 0.9533 / 0.9921 & 0.8247 / 0.9215 & 0.8416 / 0.9302 \\
Freq-only                   & 0.9587 / 0.9928 & 0.9496 / 0.9915 & 0.8089 / 0.9103 & 0.8273 / 0.9187 \\
\hline
\end{tabular}
\end{table}

\subsection*{Sensitivity Analysis}
We further analyzed the impact of two key hyperparameters on model performance: the supervised contrastive loss weight $\lambda_{\text{cont}}$ and the SFTF window size. For $\lambda_{\text{cont}}$, we evaluated values in $\{0, 0.001, 0.01, 0.1, 1.0\}$ while keeping the cross-entropy weight fixed at 1, where $\lambda_{\text{cont}} = 0$ corresponds to training with cross-entropy alone. As shown in Figure \ref{ablation}(a), both accuracy and AUC improved when a small contrastive term was introduced ($\lambda_{\text{cont}} = 0.001$) compared to the baseline. Performance peaked at $\lambda_{\text{cont}} = 0.01$, achieving an accuracy of 0.9801 and AUC of 0.9979 on CPSC2021, and 0.9507 / 0.9894 under cross-dataset evaluation. Further increasing $\lambda_{\text{cont}}$ led to slight degradation, indicating that excessive emphasis on representation learning can interfere with the primary diagnostic objective.

\begin{figure}[!htbp]
    \centering
    \includegraphics[width=0.9\linewidth]{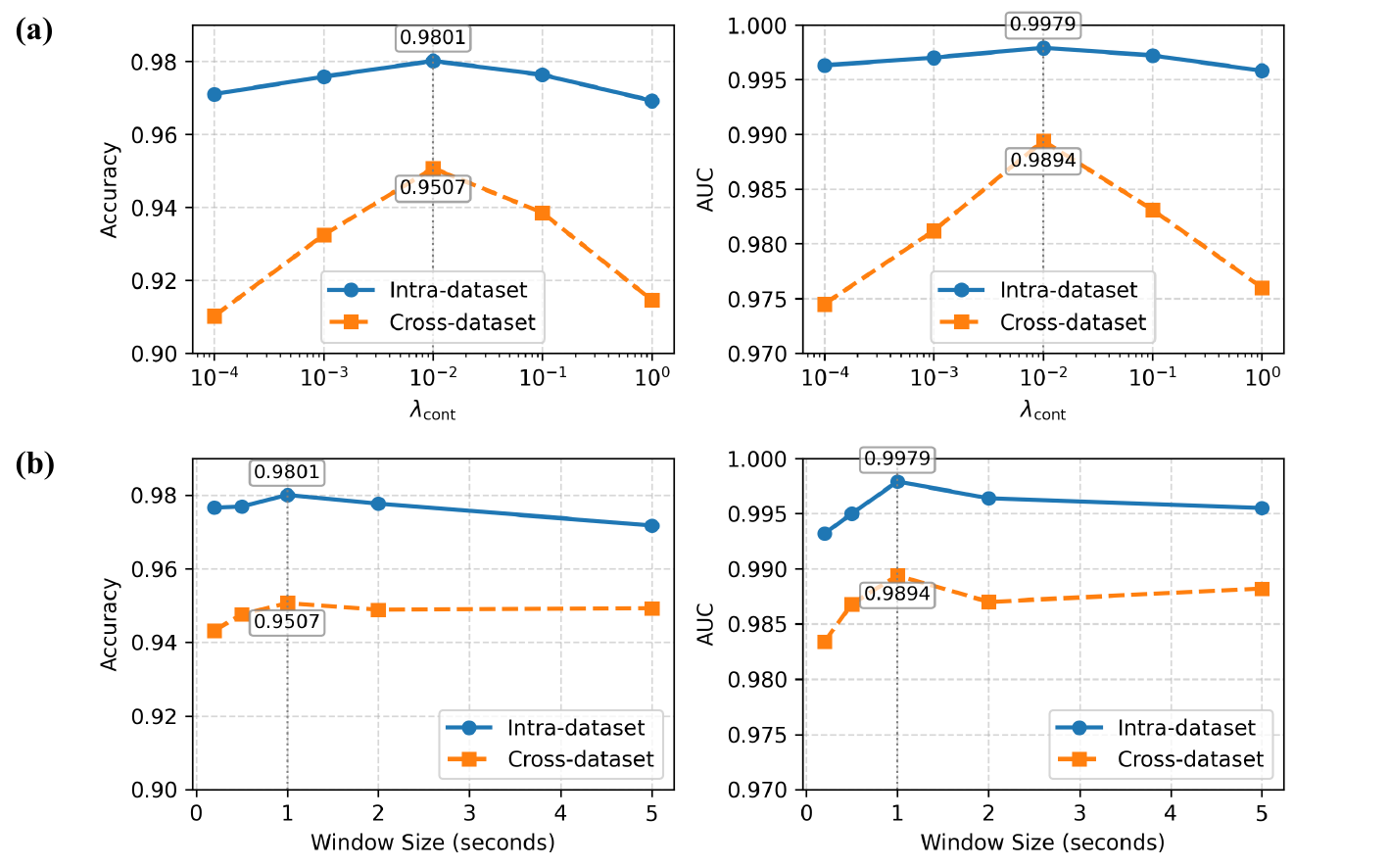}
    \caption{Accuracy and AUC versus of (a)$\lambda_{\text{cont}}$ and (b)STFT window size on CPSC2021 (intra) and CPSC2021 $\to$ AFDB (cross).}
    \label{sensitivity}
\end{figure}

Additionally, we investigated the sensitivity of the STFT module to its window size parameter, testing values of $\{0.2, 0.5, 1, 2, 5\}$ as shown in Figure \ref{sensitivity}(b). The results demonstrated remarkable stability across this wide range, with performance variations remaining marginal. The optimal performance was achieved at a window size of 1, which was subsequently adopted as the default setting. This insensitivity to window size suggests that the SFTF module captures robust temporal-frequency patterns that are not overly dependent on precise scale specification, thereby enhancing the framework's practical applicability across diverse signal characteristics. Collectively, these analyses confirm that MGCNet achieves optimal performance with $\lambda_{\text{cont}} = 0.01$ and SFTF window size = 1.

\subsection*{Interpretability}
\begin{figure}[!htbp]
    \centering
    \includegraphics[width=0.9\linewidth]{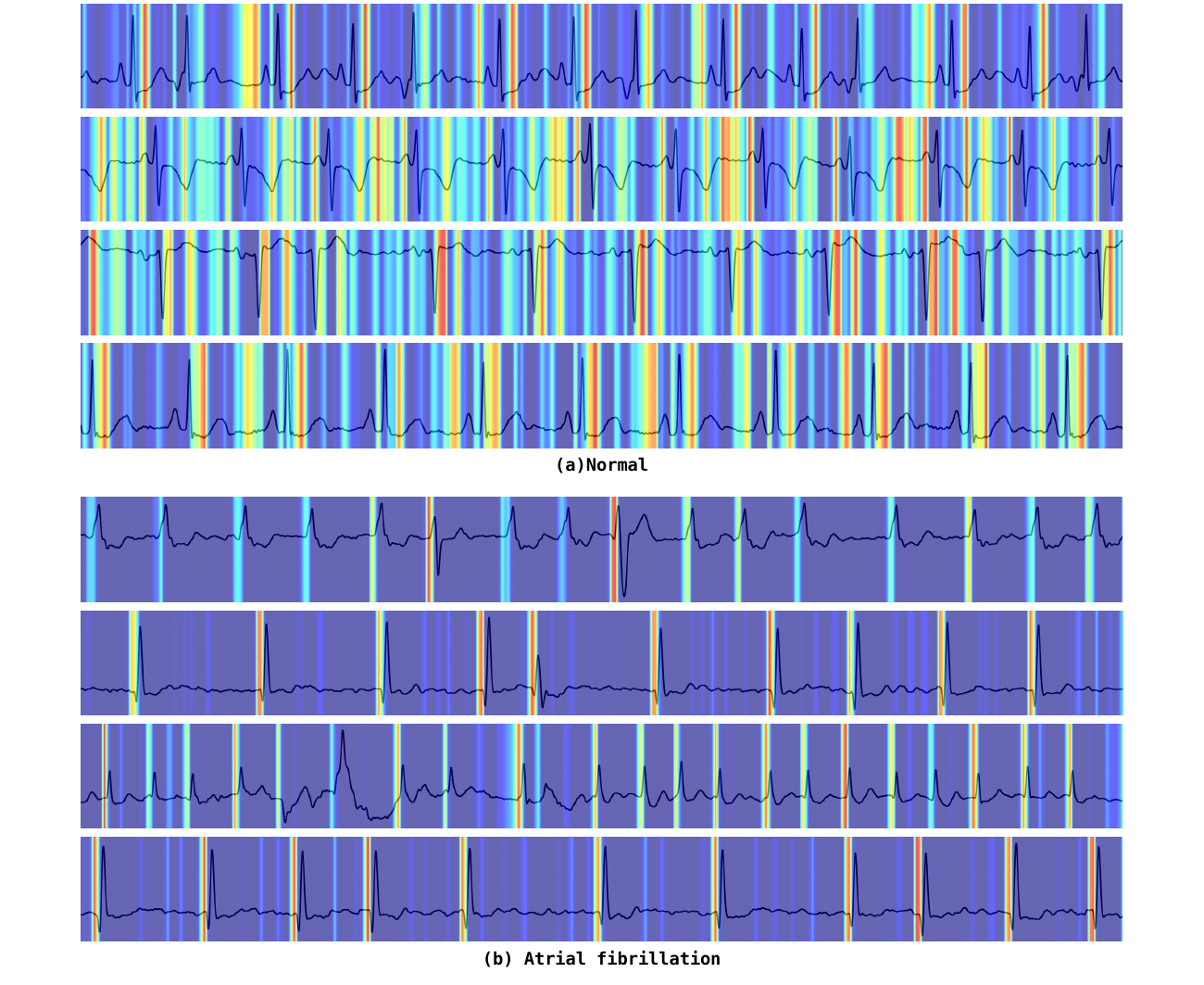}
    \caption{Grad-CAM \cite{gradcam} attention maps overlaid on ECG waveforms. Redder regions indicate higher model attention, while bluer regions indicate lower attention. Top: Normal sinus rhythm shows diffuse activation across the cardiac cycle. Bottom: Atrial fibrillation exhibits concentrated attention near the P-wave region, where organized atrial activity is typically lost.}
    \label{explain}
\end{figure}

We employed Gradient-weighted Class Activation Mapping (Grad-CAM) \cite{gradcam} on the time-domain ECG to examine which segments most influenced the model’s predictions as shown in Figure. \ref{explain}. For AF samples, Grad-CAM consistently highlighted the P-wave region—where organized atrial activity normally occurs but is typically absent or irregular in AF. This suggested the model learned to detect the loss of regular atrial depolarization, consistent with clinical AF criteria.

In normal samples, activation was more dispersed across the P-wave, QRS complex, and T-wave, with no dominant focal region. This reflected the absence of pathological markers; the model likely relied on global waveform morphology and rhythm regularity rather than a single diagnostic feature.

Grad-CAM maps were upsampled via linear interpolation from lower-resolution feature maps to match the input signal length. While this may cause mild smoothing or minor positional shifts, the regional activation trends remained robust and clinically plausible.

\subsection*{Visualization}
To provide intuitive insights into the generalization mechanism, we visualized the learned feature representations using t-distributed Stochastic Neighbor Embedding (t-SNE) \cite{tsne} dimensionality reduction, augmented with kernel density estimation (KDE) contours and confidence ellipses to quantify cluster compactness and distributional overlap. Four variants were compared: (a) w/o BGM, (b) w/o CSCL, (c) w/o both modules, and (d) the complete MGCNet. 

\begin{figure}[!htbp]
    \centering
    \includegraphics[width=0.9\linewidth]{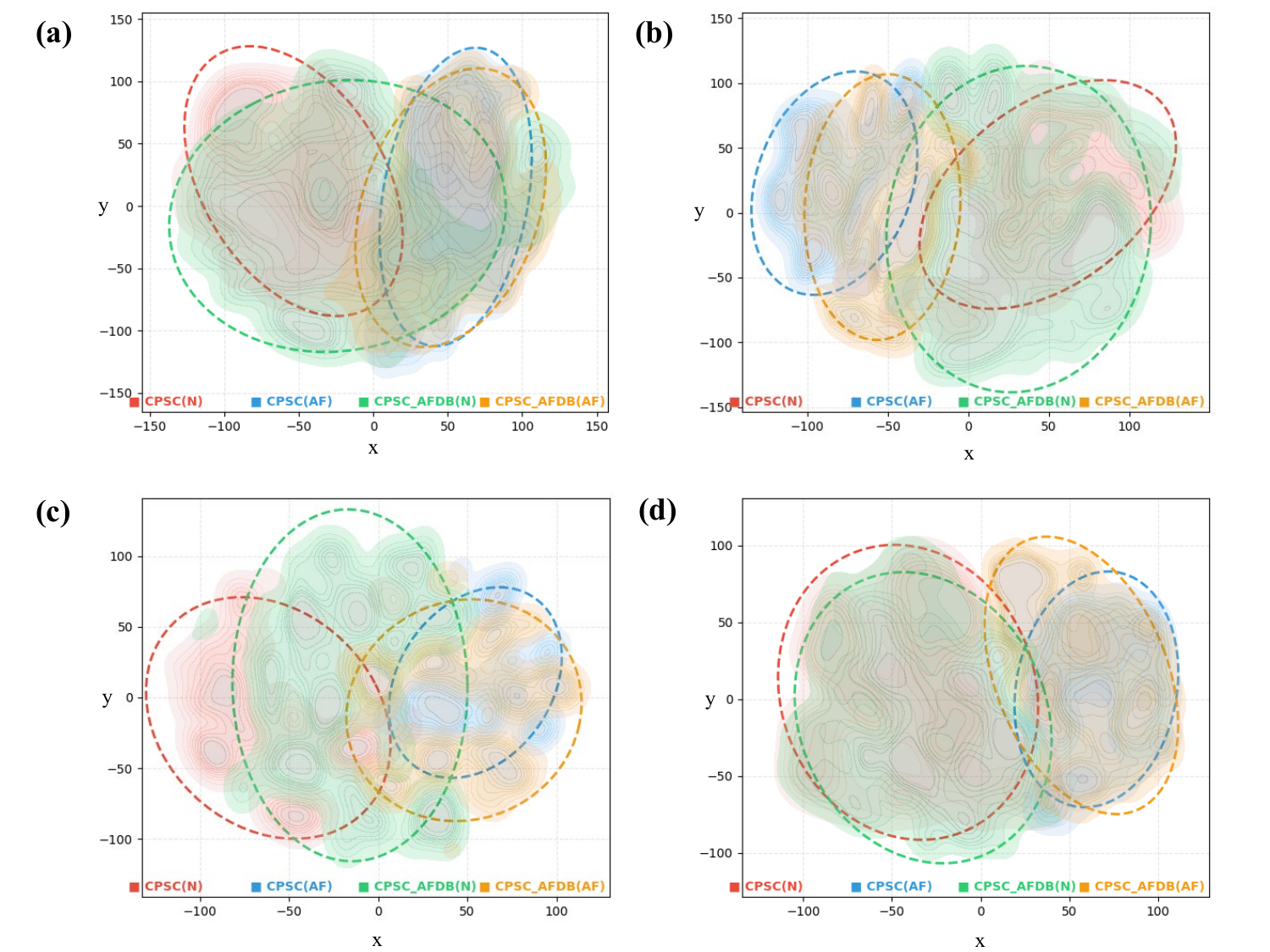}
    \caption{t-SNE \cite{tsne} visualization of feature representations extracted from the penultimate layer. The comparison evaluates four variants: (a) w/o BGM, (b) w/o CSCL, (c) w/o both modules, and (d) the complete MGCNet. "CPSC" denotes the in-domain test set (trained and tested on CPSC2021), while "CPSC\_AFDB" represents the cross-domain test set (trained on CPSC2021, tested on AFDB). In the complete MGCNet (d), the substantial overlap between "red and green" (Normal class) and "blue and yellow" (AF class) indicates that the source and target domain distributions are effectively aligned.}
    \label{tsne}
\end{figure}

As illustrated in Figure. \ref{tsne}, the baseline variant lacking both modules (c) exhibited severe domain shift, with CPSC2021-test and AFDB-test samples forming distinct, non-overlapping clusters for both AF and Normal classes. Removing either BGM (a) or CSCL (b) individually yield partial improvement, yet noticeable distributional discrepancies between the two test domains persist. In stark contrast, the complete MGCNet (d) demonstrated remarkable alignment: the KDE contours and confidence ellipses for AF and Normal samples from CPSC2021-test and AFDB-test exhibit substantial spatial overlap, indicating that the learned features were nearly indistinguishable across source and target domains. This pronounced distributional congruence confirmed that the synergistic integration of BGM and CSCL is essential for extracting multi-modal domain-invariant, physiologically grounded representations. BGM provides domain-invariant feature extraction through adaptive time-frequency selection, while CSCL enforces discriminative clustering within this invariant space, thereby directly underpinning the superior cross-dataset generalization observed in quantitative evaluations.

\begin{table}[!htbp]
\centering
\caption{Comparison of computational complexity (FLOPs) and model size (parameters) across ECG AF detection models.}
\label{complexity}
\begin{tabular}{lcccc}
\toprule
Model & FLOPs (G) & Params (M) & CPU FPS (Sample/s) & GPU FPS (Sample/s) \\
\midrule
SCCNN (2023)\cite{zshape}           & 2.07 & 2.03& 101 & 2022 \\
IMCResNet (2024)\cite{msecg1}       & 2.09 & 0.84& 50 & 588 \\
MoETransformer (2024)\cite{transecg1} & 1.12 & 2.05& 249 & 463 \\
SeqAFNet (2024)\cite{transecg2}     & 1.04 & 2.82& 76 & 710 \\
MFEGNet (2025)\cite{attenecg}       & 1.12 & 4.71& 57 & 346 \\
MSCGN (2026)\cite{msecg2}           & 0.50 & 1.38& 119 & 1136 \\
\textbf{MGCNet (Ours)}              & \textbf{0.96} & \textbf{1.96} & \textbf{64} & \textbf{867} \\
\bottomrule
\end{tabular}
\end{table}

\section*{Discussion}
\subsection*{Efficiency}
Computational efficiency was essential for deploying ECG analysis models in resource-constrained clinical or edge environments. As shown in Table \ref{complexity}, our MGCNet achieved a favorable trade-off between model size, computational cost, and inference speed. With only 0.96 G FLOPs and 1.96 M parameters, MGCNet was more lightweight than most recent AF detection models such as SCCNN, IMCResNet, SeqAFNet, and MFEGNet. In terms of inference throughput, MGCNet achieved 64 samples/s on CPU and 867 samples/s on GPU (NVIDIA GeForce RTX 3090, 24 GB VRAM; Intel Core i7-12700KF), demonstrating competitive real-time capability. While MSCGN reported the lowest FLOPs (0.50 G) and the fastest GPU inference (1136 samples/s), its smaller capacity may compromise representation ability. In contrast, MGCNet maintained sufficient expressiveness to effectively fuse time- and frequency-domain features without excessive overhead.

Combined with supervised contrastive learning, which improved feature quality without adding inference cost, our approach demonstrated that strong robustness and generalizability can be achieved alongside high efficiency, making it suitable for deployment on portable or embedded cardiac monitoring devices \cite{efficient1, efficient2}.

\begin{figure}[!htbp]
    \centering
    \includegraphics[width=0.7\linewidth]{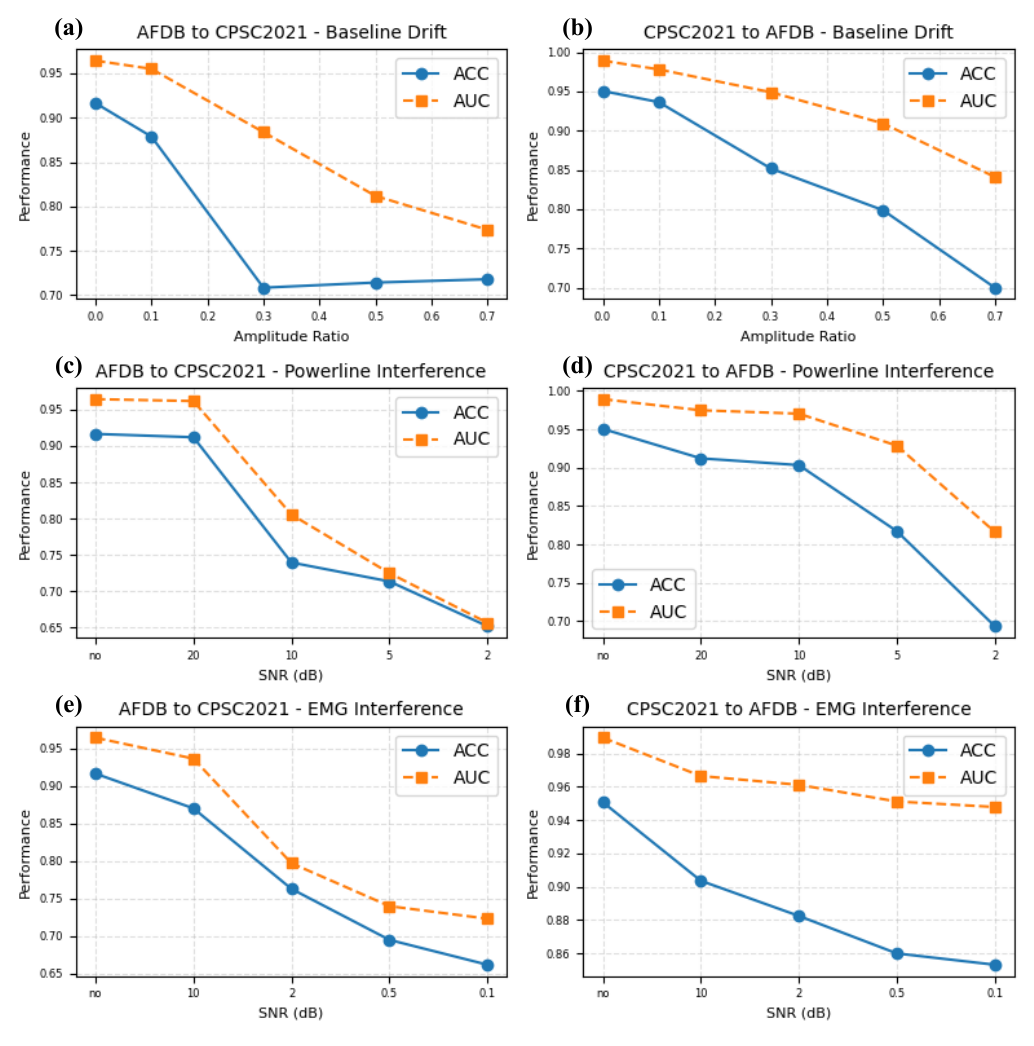}
    \caption{Robustness of cross-dataset AF classification under increasing levels of ECG interference. Left column: performance of AFDB → CPSC2021 transfer; right column: CPSC2021 → AFDB. From top to bottom: (a,b) baseline drift(0.1 Hz, amplitude ratio 0–0.7), (c,d) powerline interference (50 Hz fundamental with two harmonics, SNR = $\infty$/no to 2 dB), and (e,f) EMG interference (band-limited to 20–500 Hz, SNR = $\infty$/no to 0.1 dB). Solid lines with circles denote accuracy (ACC); dashed lines with squares denote AUC.}
    \label{noise}
\end{figure}

\subsection*{Anti-interference}
To address three common sources of degradation in real-world ECG recordings—baseline drift, powerline interference, and electromyographic (EMG) interference \cite{noise1}\cite{noise2}—we systematically evaluated the model’s anti-interference capability under two cross-dataset transfer settings: AFDB $\to$ CPSC2021 and CPSC2021 $\to$ AFDB. The result is shown in the Figure. \ref{noise}.

Baseline driftwas simulated as a low-frequency sinusoidal modulation (0.1 Hz) with increasing amplitude ratios.  Performance deteriorated significantly once the distortion exceeded a moderate level, particularly in the AFDB → CPSC2021 direction, indicating that excessive low-frequency drift can obscure critical morphological cues essential for accurate classification.

Powerline interference was introduced as a composite signal comprising a 50 Hz fundamental and its first two harmonics, with signal-to-noise ratio (SNR) progressively reduced.  The model exhibited robustness at mild interference levels but experienced pronounced performance loss under severe contamination.

EMG interference was synthesized as band-limited Gaussian noise (20–500 Hz) across a wide SNR range.  In this setting, performance declined more gradually compared to powerline interference, and the CPSC2021 $\to$ AFDB direction maintained substantially higher robustness even under extreme noise conditions.

Collectively, these findings highlighted that while the model remains vulnerable to intense low-frequency and narrowband interference, it achieved meaningful anti-interference performance, thereby supporting its applicability in practical, noise-prone cardiac monitoring environments. Notably, the CPSC2021 $\to$ AFDB transfer was consistently more robust, indicating that training on a larger and more diverse dataset improved resilience to interference.

\subsection*{Fail Cases}
\begin{figure}[!htbp]
    \centering
    \includegraphics[width=1.0\linewidth]{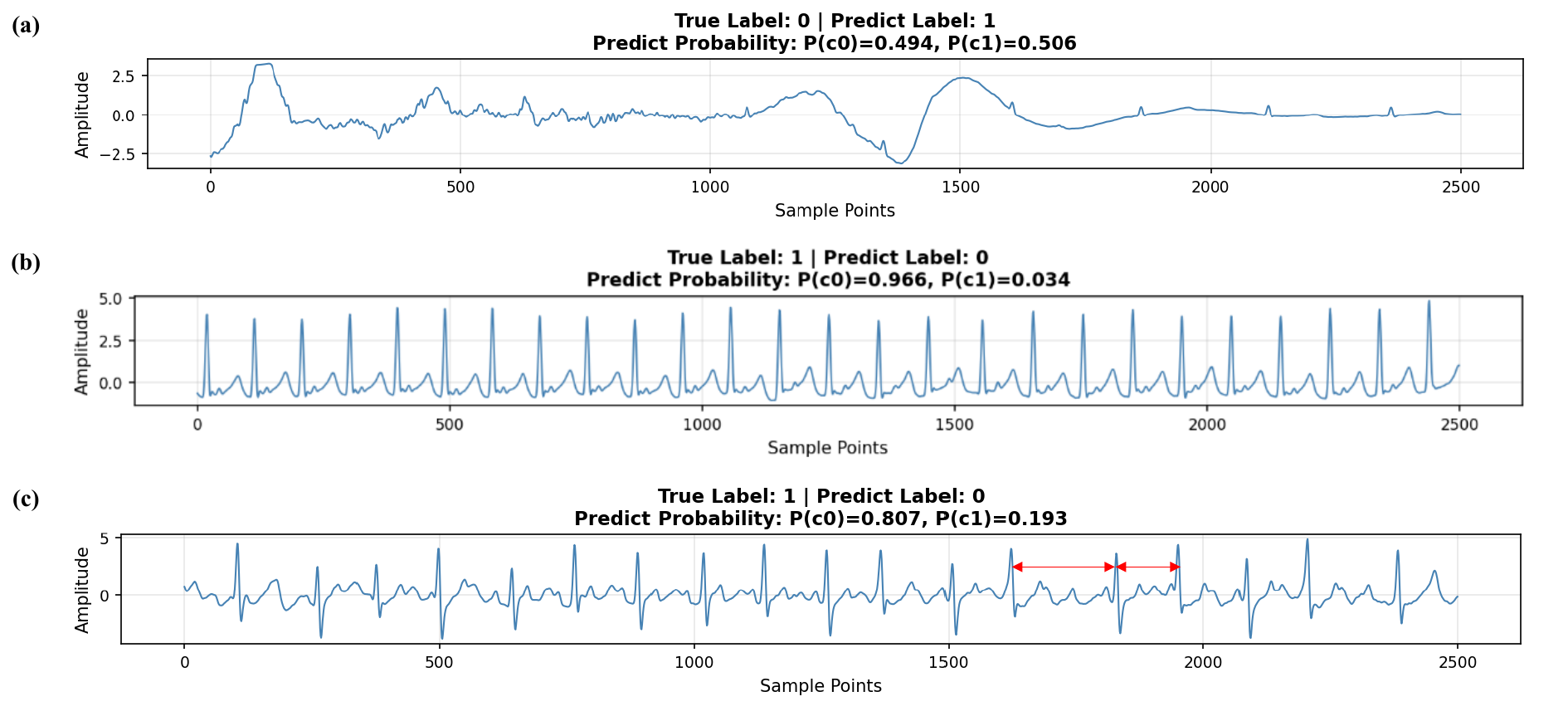}
    \caption{Failure case analysis. (a) Excessive noise leaves insufficient analyzable heartbeats for classification. (b) Ground-truth labeling error: normal sinus rhythm with distinct P-waves and regular R-R intervals mislabeled as AF. (c) Model misses transient R-R irregularities, resulting in false-negative AF detection.}
    \label{failcase}
\end{figure}

To provide a comprehensive assessment of model limitations and clinical applicability, we conducted a systematic root-cause analysis of misclassified cases. We categorized misclassified cases into three etiologies: insufficient signal quality, labeling discrepancies, and subtle features (Figure \ref{failcase}). In Figure \ref{failcase} (a), excessive noise left too few discernible heartbeats for valid classification, highlighting the need for signal quality rejection. Figure \ref{failcase} (b) shows apparent false negatives where the model correctly identified normal sinus rhythm (regular R-R intervals, distinct P-waves) despite an AF label. Figure \ref{failcase} (c) subtle features such as brief R-R irregularities that lack distinct morphological signatures, making them difficult for the model to capture. Collectively, this taxonomy revealed that failures arose from a combination of data quality issues, annotation noise, and algorithmic limitations, providing actionable insights to guide our future optimization in both preprocessing strategies and model architecture refinement.

\section*{Future Works}
Despite the promising cross-dataset generalization capability of MGCNet empowered by multimodal fusion and contrastive learning, its performance degradation under severe domain shift inevitably limits the practical deployment robustness in real-world clinical scenarios, leaving considerable room for adaptability optimization. To address this bottleneck, we propose to pursue two effective technical paradigms for enhancing the domain adaptive performance of medical intelligent models: domain adaptation (DA) and test-time adaptation (TTA).

Domain adaptation serves as an offline cross-domain alignment strategy, which leverages unlabeled target-domain data during the training phase to mitigate the distribution discrepancy between source and target domains \cite{da}. For example, Ngo et al. \cite{da2,da3} constructed a graph structural learning framework to explicitly characterize the intrinsic correlation of homogeneous samples.    On this basis, they optimized category prototype representations to enhance feature discrimination and target-domain robustness, thereby facilitating precise and effective cross-domain knowledge transfer for medical vision tasks.

Different from offline domain adaptation, test-time adaptation is a dynamic, online adaptive paradigm. It enables pre-trained fixed models to implement real-time parameter fine-tuning during the inference stage \cite{tta}. For example, Chen et al. \cite{tta2} validated the efficacy of this paradigm via a dual-branch collaborative optimization strategy integrating contrastive learning and online pseudo-label refinement. Specifically, contrastive learning promotes the model to learn domain-shift-invariant feature representations, while high-quality dynamically optimized pseudo-labels further inject semantic prior constraints into the contrastive learning process, realizing efficient and robust test-time domain adaptation.

Collectively, the complementary combination of offline cross-domain distribution alignment and online inference-stage self-adaptation provides a feasible and systematic technical roadmap for the development of robust, generalizable, and clinically adaptable next-generation medical ECG artificial intelligence systems.

\section*{Conclusions}
We presented a robust and generalizable deep learning framework for atrial fibrillation (AF) detection from single-lead ECG signals, combining time- and frequency-domain features through a dedicated fusion architecture.   By incorporating supervised contrastive learning, the model learns discriminative embeddings that enhance intra-class compactness and inter-class separation, significantly improving cross-dataset generalization. 
Experiments across multiple public datasets demonstrate that our approach achieves high accuracy while maintaining strong performance under domain shifts.  Post-hoc Grad-CAM analysis further confirms that the model focuses on clinically relevant regions, particularly the P-wave interval in AF cases, where organized atrial activity is typically lost.
These results indicate that integrating multi-domain representation with contrastive representation learning is an effective strategy for building reliable, interpretable, and deployable ECG classification systems.   Future work will focus on extending this framework to multi-class arrhythmia detection and optimizing it for resource-constrained clinical environments. Additionally, we aim to integrate domain adaptation and test-time adaptation strategies to ensure robust performance against distribution shifts in real-world scenarios.

\section*{Acknowledgements}
During the preparation of this manuscript, the author(s) used Qwen (Alibaba Cloud, Qwen-Max, 2025) for the purposes of language polishing, improving the clarity and flow of the text, and assisting in the drafting of methodological descriptions.  The authors have reviewed and edited all AI-generated content and take full responsibility for the accuracy and integrity of the final manuscript.

\section*{Author contributions statement}
Conceptualization, H.L. ,J.W.    and X.O.;      methodology, H.L. Y.L.     and X.O.;      software, X.O;      validation, Y.L. and X.O.;      formal analysis, H.L., J.W., J.X., M.L., S.L., and X.O.;      investigation, X.O.;      resources, H.L. and J.X.;      data curation, X.O.;      writing---original draft preparation, X.O.;      writing---review and editing, H.L., Y.L. and J.W.;      visualization, X.O.;      supervision, H.L. and Y.L.;      project administration, H.L.,J.X.,Y.L. and X.O.;      funding acquisition, H.L. All authors have read and agreed to the published version of the manuscript.

\section*{Competing interests}
The authors declare no conflicts of interest.

\section*{Data availability}
The AFDB dataset is publicly available from PhysioNet (\url{https://www.physionet.org/content/afdb/1.0.0/}). The CPSC2021 dataset was provided by the China Physiological Signal Challenge 2021 and is accessible via the official challenge website (\url{https://physionet.org/content/cpsc2021/1.0.0/}). The LTAF dataset is publicly available from PhysioNet (\url{https://physionet.org/content/ltafdb/1.0.0/}). The SHDBAF dataset is publicly available from PhysioNet (\url{https://physionet.org/content/shdb-af/1.0.0/}). 

\section*{Code availability}
The code and trained models are publicly available at \url{https://github.com/Ou-Young-1999/MGCNet} and archived on Zenodo with DOI: \href{https://doi.org/10.5281/zenodo.21070128}{10.5281/zenodo.21070128}.

\section*{Informed consent}
Not applicable.

\section*{Funding}
This research received no external funding.

\bibliography{references}

\end{document}